\documentclass[%
 reprint,
superscriptaddress,
 aps,
]{revtex4-2}

\usepackage{hyperref}
\usepackage{multirow}       
\usepackage[caption=false]{subfig}
\usepackage{graphicx}
\usepackage{amsmath}        
\usepackage{amssymb}        
\usepackage{physics}        
\usepackage{xcolor}
\usepackage{xcolor}

\newcommand{\ns}{\mathrm{~ns}}       

\newcommand{\MHz}{\mathrm{~MHz}}     
\newcommand{\GHz}{\mathrm{~GHz}}     
\newcommand{\sigvec}{\vec{\sigma}}

\usepackage{soul}   

\newcommand{\VTPHYSICS}{Department of Physics, Virginia Tech, Blacksburg, VA 24061}
\newcommand{\VTQUANTUM}{Virginia Tech Center for Quantum Information Science and Engineering, Blacksburg, VA 24061, USA}

\begin{document}

\preprint{APS/123-QED}


\title{Implementing and benchmarking dynamically corrected gates on superconducting devices using space curve quantum control}

\author{Hisham Amer}
\author{Evangelos Piliouras}
\author{Edwin Barnes}
\author{Sophia E. Economou}
\affiliation{\VTPHYSICS}
\affiliation{\VTQUANTUM}


\begin{abstract}
We use Space Curve Quantum Control (SCQC) to design, experimentally demonstrate, and benchmark dynamically corrected single-qubit gates on IBM hardware, comparing their performance to that of the standard gates provided by IBM. Our gates are designed to dynamically suppress both detuning and pulse-amplitude noise, with gate times as short as 88 ns. We compare our gates against those of IBM on two separate IBM devices and across sets of up to 18 qubits. Randomized benchmarking is done utilizing our detuning- and amplitude-robust gates in randomized Clifford circuits containing up to 4000 gates. Our gates achieve error-per-Clifford rates that reach as low as 7$\times10^{-5}$ ($\pm10^{-6}$) and which remain nearly constant as the compound noise is increased up to 4\% amplitude noise and up to a detuning noise of 342 kHz; this is in contrast to the IBM gates, which exhibit rates that drop to order $10^{-3}$ across this range. This range is consistent with the commonly reported frequency fluctuations and with the upper bound of the statistical uncertainty in gate calibration. In addition, we investigate the performance across larger noise ranges of up to 20\% amplitude and 3.5 MHz detuning noise using quantum process tomography. Finally, we experimentally demonstrate how SCQC can be tailored to different practical use cases by trading off amplitude-robustness for ultrafast 60 ns dephasing-only robust pulses. Our work establishes experimental guidelines for implementing SCQC-designed dynamically corrected gates on a broad range of qubit hardware to limit the effect of noise-induced errors and decoherence.
\end{abstract}

\maketitle


\section{Introduction}
\label{intro}
Practical universal gate-based quantum computing requires precise and accurate control over the evolution of a quantum register in a noisy environment. Despite the steady improvement in the performance of physical gates, much work is still needed to reliably utilize fault tolerant QEC approaches, in noisy environments, below thresholds needed for large-scale practical quantum computing \cite{acharya2024_QEC_google_subthreshold}. This demand also extends to the recent non-fault tolerant quantum utility approaches \cite{Kim2023, IBMQuantum}.

Suppressing the effects of noise that causes stochastic parameter shifts in the device and control Hamiltonians is particularly important as these are typically key sources of gate infidelity and decoherence. On superconducting devices, these shifts can occur on different timescales \cite{Tripathi_2024, Burnett_2019_study_dephasing_with_DD} and are most commonly described in terms of low-frequency $1/f$ noise \cite{1/f_0.58-0.8_1987,
Rower_2023_1/f_supression,
Christensen_2019_1/f_1.9,
1/f_review_RevModPhys.86.361,
Krantz_2019_engineer_guide,
Tripathi_2024,
Burnett_2019_study_dephasing_with_DD,
Kumar_2016_1/f_flux_origin}
that arises from two-level charge fluctuators in the surrounding material, flux noise, ionizing radiation, and quasiparticles~\cite{
Tripathi_2024,
Krantz_2019_engineer_guide,
Burnett_2019_study_dephasing_with_DD,
Burnett_2014,
Kumar_2016_1/f_flux_origin,
Houck_2009,
Klimov_2018_tls_qubit_fluctuations, decoherence_IBM_TLS_PRL, 
high-energy_impacts_TLS_google, Thorbeck_2023_TLS_impact_IBM,
Barrett_2023_freq_calibration_error}. While these shifts in device parameters can necessitate frequent gate re-calibrations, calibration processes themselves exhibit inherent uncertainties related to their experimental nature, adding further to calibration and device parameter mismatches. That said, for the gates to consistently operate with high fidelity, not only do the calibrated gate parameters, with their inherent statistical uncertainty, need to fall close enough to their ideal values, but the calibrated gates should also remain faithful to their target operation throughout the widely varying experimental run-times. Accordingly, fluctuations in the device and control parameters create the persistent need for gate re-calibrations, significantly adding to the experimental overhead. This re-calibration step is typically a necessary routine that is performed prior to each experiment conducted in most labs and is especially important for cloud devices, such as the IBM superconducting devices used in this work. Such system-wide calibration data is available to authorized users and gets updated several times a day \cite{IBMQuantum}. An example of how this re-calibration step can interrupt the experimental workflow in real time can be seen in recent below-threshold QEC surface code demonstrations carried out by groups like Google Quantum \cite{acharya2024_QEC_google_subthreshold}. Namely, in these QEC demonstrations, large-scale gate re-calibrations were needed to maintain favorable physical qubit error rates in the face of unavoidable parameter drifts, especially in lower-distance code implementations. 

Regardless, even when the calibration overhead is accounted for, there are still down times in between where the gates are still vulnerable to noise-induced errors which can happen in between experiments or within the same experiment. Additionally, the control field calibration process itself involves real measurements and curve fitting of experimental data, and therefore is associated with some statistical uncertainties; a detail that is overlooked in most, if not all, gate-robustness studies. The latter either tend to overshoot the noise resolution necessary to investigate this unavoidable error source, or they employ benchmarking experiments that lack the precision necessary to resolve these finer errors. The range of uncertainty in frequency calibrations reported by others, and also found in the present work, is on the order of 10's to 100's of kHz~\cite{IBMQuantum,
Kanazawa2023_Qiskit_exp, Barrett_2023_freq_calibration_error}, further justifying the need for noise-robust control. 

There are several ways to deal with noise in a quantum device, for example,  through surface treatments in superconducting circuits \cite{de_Graaf_2018}, or more drastic hardware design changes such as the addition of a large shunting capacitance to charge qubits to reduce their charge sensitivity, as in transmon qubits~\cite{Koch_2007,Houck_2009}. A complementary way to improve noise-robustness is to design it directly into the control fields that generate the gates. Dynamically corrected gates (DCGs) \cite{ GOELMAN1989423,
Viola1998,
Biercuk2009,
Khodjasteh2009,
Khodjasteh2010,
Wang2012,
DCGGateDecoherence,
Khodjasteh2012,
Kestner2013,
Green2013,
DCG_Lie_KBrown,
Barnes2015,
DCGCNOT,
Buterakos2018,
Kestner_2022_DCG_NN,
UtkanDCG2qub,
Kanaar2021,
Carvalho2021,
Zeng2019,
Dong2021,
Buterakos2021,
Li2021,
Barnes2022,
Zhuang2022,
Nelson2023,
piliouras2025,
qurveros,
PhysRevLett.132.250604SCQC, walelign2024dynamicallycorrectedgatessilicon,
Poggi_2024} are typically designed to achieve a target gate operation through a favorable choice of control fields that suppress noise-induced errors and decoherence. Accordingly, DCGs must achieve two objectives: executing the target gate correctly in the absence of noise, or \textit{gate-fixing}, and suppressing noise errors during the gate, i.e., \textit{noise-robustness}.

Numerically generated DCGs rely on parameterizing the control field followed by a numerical search for locally optimal solutions; this search is powered by algorithms including different versions of GRAPE \cite{Khaneja2005GRAPE}, CRAB \cite{Doria_2011_CRAB}, CORPSE \cite{Jones2012CORPSE}, GOAT \cite{GOAT_PhysRevLett.120.150401} or even machine learning approaches \cite{Kestner_2022_DCG_NN}. Consequently, this locality translates to an optimization over control field parameters that inextricably combines the gate-fixing and noise-robustness conditions together in the same cost function, which in practice typically leads to some form of trade-off between these two objectives, despite the fact that such a trade-off is in principle avoidable~\cite{Poggi_2024}. On the other hand, partially or fully analytical approaches have been under-utilized in experiments~\cite{Hansen_APR2022,PhysRevLett.132.250604SCQC, walelign2024dynamicallycorrectedgatessilicon} despite offering advantages in  optimizability and experimental feasibility.

A geometrically inspired, largely analytical approach to DCGs that offers many of the aforementioned benefits is Space Curve Quantum Control (SCQC)~\cite{
Zeng2019,
Dong2021,
Buterakos2021,
Li2021,
Barnes2022,
Zhuang2022,
Nelson2023,
piliouras2025,
qurveros}. SCQC maps quantum evolution to space curves in Euclidean space in such a way that control fields are encoded in geometric properties of the space curves. Most importantly, noise-robustness conditions are mapped to simple geometric conditions that must be imposed on the space curves. The SCQC approach is practical for experiments, as it delivers entire solution sets of robust control pulses, leaving it up to the user to select the pulses that best suit their experimental setup.

In this paper, we experimentally demonstrate and benchmark the effectiveness of SCQC-designed gates  by showing that they can outperform standard IBM gates on IBM devices. Our work employs the automated space curve generation method known as B\'ezier Ansatz for Robust Quantum (BARQ) control \cite{piliouras2025}, which is available in the python package \texttt{qurveros} \cite{qurveros}. BARQ leverages the SCQC formalism such that gate-fixing and dephasing-robustness are analytically built into the space curve ansatz upfront, while other experimentally desirable control field properties are independently achieved via curve optimization. We use BARQ to design single-qubit gates that are simultaneously robust to both dephasing noise and pulse-amplitude noise. We experimentally demonstrate that our BARQ-designed gates exhibit a clear advantage over the standard IBM gates in the presence of the noise fluctuations typically measured on superconducting devices, including noise from statistical uncertainty in the calibration. The benchmarking we perform here has the precision needed to resolve gate error differences at the $10^{-5}$ level. While we choose to use superconducting devices, specifically IBM devices, for our demonstration in this work, it is important to emphasize that SCQC can be applied to any type of hardware (see Refs.~\cite{Hansen_APR2022, walelign2024dynamicallycorrectedgatessilicon} for recent applications in Si spin qubits), and the methodologies presented here can be translated to other qubit platforms.

This paper is arranged as follows. In Section \ref{sec:pulse_generation}, we cover the relevant magnitudes of the noise from the sources covered in the introduction; this is followed by a summary of how we use the SCQC formalism to generate robust gates. Section \ref{sec:experimental_methods} describes our experiments, including results for a Hadamard gate, an $X$ gate, and a $\sqrt{X}$ gate, thereby covering a subset of the Cliffords whilst also allowing us to make use of the fact that one can express any single-qubit unitary using two $\sqrt{X}$ and three appropriately angled virtual $Z$ gates \cite{McKay_2017}. Initially, all the gates will be robust to both frequency and amplitude errors. However, in Section~\ref{sec::single_robust}, we demonstrate the versatility of the SCQC approach through a practical use-case where we loosen the robustness conditions towards a single error source in exchange for shorter singly-robust pulses. Finally in Section~\ref{sec:utility_scale} the control scheme is automated to 18 qubits on \textit{ibm\_strasbourg} to investigate how well it performs over a wide range of realistic qubit conditions while we push our gates towards their coherence limit \cite{
coherence_1_Sundaresan_2020,
coherence_2_Garion_2021,
coherence_3_Wei_2024}.

\section{Robust pulses for multiple noise sources}
\label{sec:pulse_generation}

\subsection{Noisy device and control parameters}
\label{sec::pulse_generation::noise_intervals}
We start by considering a noiseless two-level approximation of the transmon Hamiltonian in the the drive frame and under the rotating wave approximation (RWA):
\begin{equation}
    \label{eqn:H_0_text}
    H_{0}(t)=\frac{\Delta}{2} \sigma_z
    +\frac{\Omega(t)}{2}\left[\cos \Phi(t) \sigma_x+\sin \Phi(t) \sigma_y\right],
\end{equation}
where {$\sigma_i$} are the Pauli matrices. The driving field has amplitude $\Omega(t)$, phase $\Phi(t)$, and detuning $\Delta$.

The Hamiltonian in Eq.~\eqref{eqn:H_0_text} can be subjected to multiple sources of noise or miscalibrations. Common superconducting device noise sources can lead to unwanted fluctuations in the qubit energy levels typically on the order of 10s of kHz with reports of up to 500 kHz, with larger values being more common in flux tunable qubits~\cite{Burnett_2019_study_dephasing_with_DD, Klimov_2018_tls_qubit_fluctuations, decoherence_IBM_TLS_PRL, 
Barrett_2023_freq_calibration_error}. Some research groups have also reported frequency shifts of up to a few MHz on superconducting processors \cite{high-energy_impacts_TLS_google,Thorbeck_2023_TLS_impact_IBM}, which are thought to result from exposure to ionizing radiation, one of the main sources of errors in large-scale QEC experiments \cite{acharya2024_QEC_google_subthreshold}. Lastly, noise affecting the pulse waveform can also be observed due to imperfect control hardware.

A noise model that applies to a wide range of qubit platforms, including superconducting devices, involves additive longitudinal fluctuations $\delta_z$ in the detuning caused by energy level fluctuations or driving frequency miscalibrations. Additionally, it may also involve multiplicative driving amplitude fluctuations $\epsilon$. These noise sources can be added to Eq. \eqref{eqn:H_0_text} by taking
\begin{align}
    \Omega(t)&\to (1+\epsilon)\Omega(t) ,\label{multiplicative_Rabi} \\
    \Delta &\to \Delta +\delta_z \label{additive_dephasing}.
\end{align}

\subsection{Robust pulse generation with SCQC}\label{sec::pulse_generation::summary_BARQ_SCQC}
The generation of control pulses that take into account the aforementioned control parameter fluctuations is not a straightforward task. To understand why, we can express the quantum evolution $U$ as a product of two contributions; the noise-free (ideal) evolution $U_0$ and the noisy part $U_{\text{noise}}$ with $U = U_0U_{\text{noise}}$. The noise-free dynamics are governed by $U_0$, as it is the solution to the Schr\"{o}dinger equation $i\dot U_0 = H_0U_0$ achieving a desired gate $U_g$ at a final time $T_g$, i.e., $U_0(T_g)=U_g$. However, if one hopes to suppress the effects of noise, the design procedure must also aim to enforce $U_{\text{noise}}(T_g) \approx I$, where $I$ is the identity.
As shown in Appendix~\ref{app::theory_BARQ}, we can use the Magnus expansion of $U_{\text{noise}}$ to first-order in the noisy parameters to yield the first-order robustness conditions:
\begin{align}
    \delta_z:\int_0^{T_g} dt& \, U_0^\dagger \sigma_z U_0 = 0 ,\label{eqn:dephasing_robust_cond_txt}\\
     \epsilon:\int_0^{T_g} dt&\, U_0^\dagger\, \Omega(t)\left[\cos \Phi(t) \sigma_x+\sin \Phi(t) \sigma_y\right]U_0 =0 .\label{eqn:amp_roubst_cond_txt}
\end{align}
SCQC then makes the connection to geometric space curves by defining the curve  $\vec{r}(t)$ through \cite{Zeng2019}
\begin{align}
    \int_0^{t} dt' \, U_0^\dagger \sigma_z U_0 = \vec r (t) \cdot \sigvec. \label{scqc_main_eq}
\end{align}
If $\vec r(t)$ is taken to be the position vector defining a space curve in three dimensions, then the $(\epsilon,\delta_z)$ robustness conditions in Eqs.~\eqref{eqn:dephasing_robust_cond_txt},\eqref{eqn:amp_roubst_cond_txt} become the geometric space curve conditions given in Eqs.~\eqref{eqn:closed_curve_geometry_app},\eqref{eqn:zero_area_geometry_app}, which describe a closed space curve whose tangent vector traces out zero oriented area; we refer to these as the \textit{closed-curve} and \textit{zero-area} conditions respectively; these conditions can be satisfied exactly analytically~\cite{Nelson2023}. SCQC similarly allows higher-order robustness conditions, and robustness to other noise sources, to be mapped to geometric analytical conditions~\cite{Barnes2022}. 

The full characterization of the quantum evolution, as governed by the Schr\"{o}dinger equation, can then be recast in terms of differential geometry and the Frenet-Serret (FS) equations that govern the shapes of space curves. Once a space curve with the desired properties is constructed, the control fields $\{\Omega(t),\Phi(t),\Delta(t)\}$ can be extracted from the space curve's curvature $\kappa(t)$ and torsion $\tau(t)$ (see Appendix~\ref{app::theory_BARQ}).

The immense freedom in finding admissible noise-robust control waveforms can now be understood as a natural geometrical fact, since many different space curves can satisfy the geometrical constraints corresponding to the gate-fixing and noise-robustness conditions. To generate noise-robust gates in an automated fashion, we employ a protocol some of us recently developed termed BARQ~\cite{piliouras2025}. In BARQ, the position vector $\vec{r}$ is parameterized using a set of control-points and basis-functions based on the Bernstein basis (see Appendix~\ref{app::theory_BARQ}). A subset of control points is responsible for setting the space curve boundary conditions that both fix the gate operation $U_g$ to unit fidelity in the noise-free case, and satisfy the closed-curve condition that guarantees robustness against detuning noise. The remaining control points can then be optimized to achieve other desired properties such as robustness to amplitude noise, without introducing any trade-off between gate-fixing and noise-robustness, as the noise-free gate is always guaranteed to have unit fidelity, and noise-robustness can be incorporated independently of this fact. This separation between the two DCG objectives means that the designed pulses can usually be run on devices straight out of the simulator with minimal to no calibration, as demonstrated experimentally in Section~\ref{sec:experimental_methods}. The curve optimization and pulse generation are handled by the Python package \texttt{qurveros}~\cite{qurveros}.

\section{Experimental Methods of SCQC on real devices}
\label{sec:experimental_methods}

To benchmark robustness, we evaluate gate performance in two ways. First, quantum process tomography (QPT) is used to reconstruct the gate as a quantum channel, specifically the Choi matrix, which captures the the gate's response to different noise sources. To get the most out of QPT, we scan a wide range of noise strengths, which allows us to accomplish two goals: (i) We can show that the noise-robustness of our gates extends to very high levels of noise, including the radiation-induced MHz ranges discussed in Section~\ref{sec::pulse_generation::noise_intervals}. (ii) The contrast between the noise-free and strong-noise Choi matrices sheds light on the mechanism with which the noise affects the final gate. The QPT analysis however has drawbacks, including sensitivity to state preparation and measurement (SPAM) errors and its relatively wide statistical uncertainty, limiting its ability to reliably discern between fidelities above the 0.999 level. Consequently, we need a second round of analysis where we focus on the sub-500 kHz noise window where we expect most of the parameter fluctuations and calibration statistical uncertainty regions to lie. For this purpose, we use randomized benchmarking (RB) as our second benchmarking tool in order to resolve gate errors in the $10^{-4}$ to $10^{-5}$ range.

\subsection{Tailoring the pulses to qubits}
\label{sec::tailoring}

With BARQ, since the gate-fixing and detuning-noise-robustness conditions are ensured upfront by setting the control points near the ends of the space curve appropriately, we can optimize the remaining control points controlling the interior of the space curve to satisfy additional constraints for other properties such as amplitude-noise-robustness, how far off resonance the drive can be, how robust the gate is to second-order detuning noise, or how fast the rise times are. Each additional property is included as a term in the cost function used by the optimizer, and different weights can be chosen according to the relative importance of each property. Once the curve optimization is completed and the control fields are extracted form the space curve, the user can scale the pulse to specific qubit parameters. In this work, we utilize IBM's $127$-qubit Eagle transmon processors, which also included access to pulse level control with Qiskit Pulse \cite{Alexander_2020_Qsikit_pulse}. 

To scale the pulse to a given qubit on a real device, we need to specify a maximum Rabi rate $\Omega_{\text{Max}}^{\text{IBM}}$, which in our case will be on the order of tens of MHz, while the gate time $T_g$ should be on the order of tens of nanoseconds. There is an implicit relation between how strong and how fast a pulse can be whilst still having it generate the same unitary operation, and that can be captured by the unitless quantity $T_g\Omega_{\text{Max}}$, which should remain constant for any control field scaling or pulse stretching action we perform on a given control pulse. This fact is captured by the relation
\begin{equation}
    \label{eqn:Scale_Omega}
    T_g^{\text{SCQC}} \Omega_{\text{Max}}^{\text{SCQC}}
    = T_g^{\text{IBM}} \Omega_{\text{Max}}^{\text{IBM}}.
\end{equation}
This correspondence ensures that when stretching or shrinking the duration of pulses, we preserve the relation between the real and imaginary parts, whilst also preserving the area underneath the pulse envelopes, which would ideally preserve the dynamics.  

Now in Eq.~\eqref{eqn:Scale_Omega}, given that $T_g^{\text{SCQC}}$ is chosen to be unity by design, we are left with the freedom to stretch the pulse on the device, as long as our choice of $T_g^{\text{IBM}}$ and $\Omega_{\text{Max}}^{\text{IBM}}$ equals $\Omega_{\text{Max}}^{\text{SCQC}}$. This freedom to stretch the pulse gives us a lot of leeway in choosing gate times $T_g$ on a given qubit. The question of how fast the gate should be will be explored in Section~\ref{sec:utility_scale}, when we apply this on a utility scale as we try to get multiple qubits to operate at the coherence limit for some $T_g$, balancing the detrimental effects of leakage from shorter pulses against the larger impact of decoherence seen with longer pulses.

\subsection{Experimental benchmarking: doubly-robust Hadamard gate}
\label{sec::Hadamard}

\subsubsection{Hadamard QPT Setup and Motivation}
\label{sec::Hadamard::setup}
\begin{figure*}
    \subfloat{
        \includegraphics[width=0.95\textwidth]{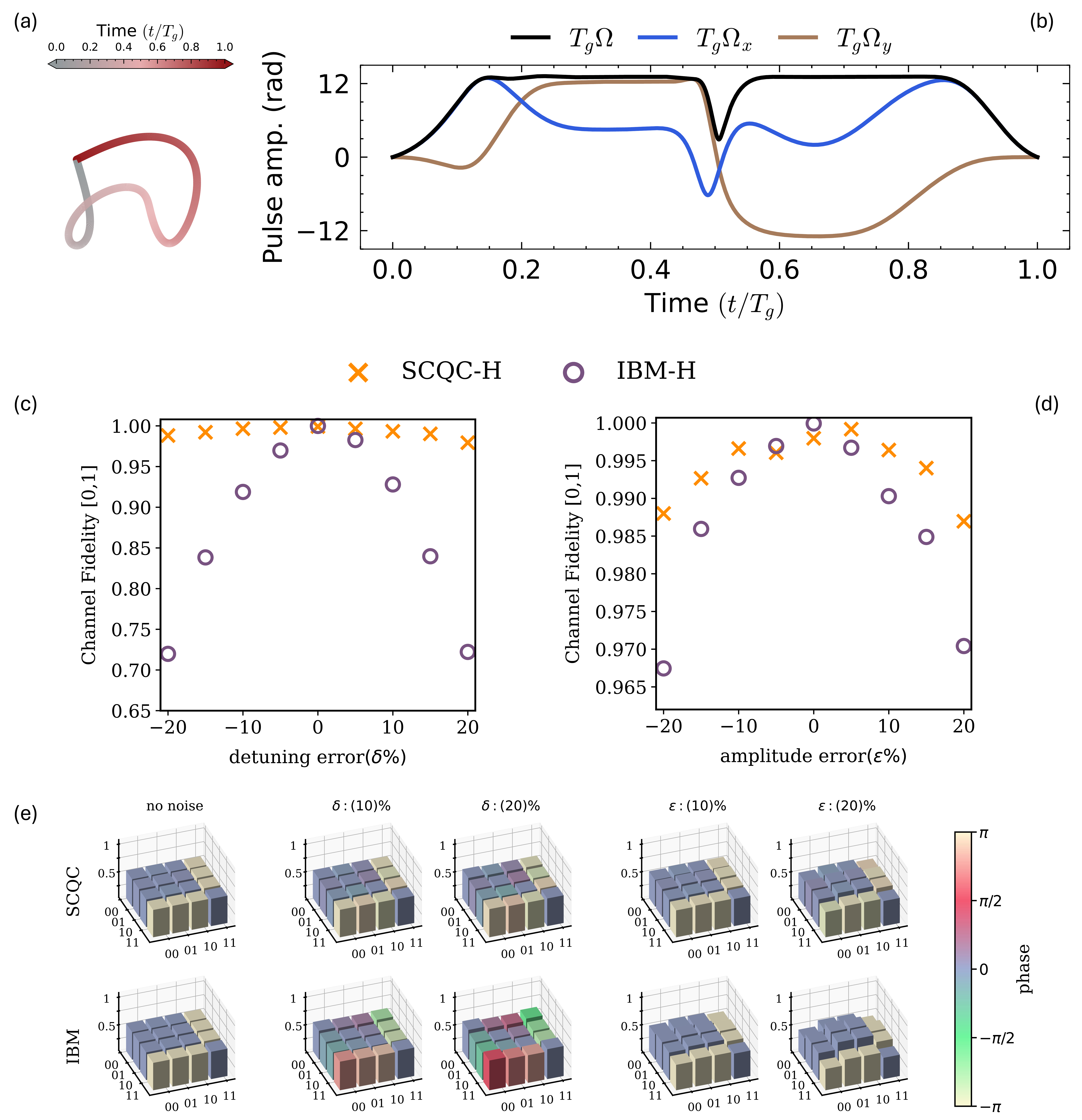}
    }
    \hfill

    \caption{(a) Depiction of the space curve generated by BARQ for a doubly-robust Hadamard, with a color bar to indicate how the space curve is traversed. The space curve satisfies the \textit{closed-curve condition} which translates to first-order robustness to detuning errors. (b) The pulse waveform extracted from the space curve.
    (c, d) Experimental results on qubit[16] of \textit{ibm\_strasbourg} for the detuning and amplitude robustness tests, respectively. The results show that SCQC pulses are considerably more noise robust than standard IBM pulses. (e) Experimental Choi matrix for $(\epsilon, \delta_z)$ = [0,10,20]\%.}
    \label{fig::hada}
\end{figure*}

For our first demonstration we implement a Hadamard gate. In this specific run of BARQ, we prioritize the control pulses being on resonance, short, smooth with a reasonable rise time and finally doubly robust, meaning robust to both dephasing and amplitude noise to first order in the noise parameters $\epsilon$ and $\delta_z$, respectively, as modeled by Eq.~\eqref{eqn:H_noise}. This Hadamard is fairly similar in its setup to that of the pulse adopted in the original BARQ paper~\cite{piliouras2025}. The final Hadamard pulse is shown in Fig.~\ref{fig::hada}(b), along with its SCQC space curve in Fig.~\ref{fig::hada}(a).

Due to the finite time available on the IBM devices, we only performed QPT for the Hadamard gates due to their implementation with a rotation axis beyond the single $X$, $Y$, or $Z$ axis. By reconstructing the full Choi matrix, we closely studied how dephasing and amplitude noise affected the SCQC gates versus the IBM gates. Fidelities were calculated using the channel fidelity, which is simply the state fidelity between the ideal and the experimentally reconstructed Choi matrices.

For the benchmarking we utilize the wide interval $[-20, 15, -10, -5, 0, 5, 10, 15, 20]\%$ for both $\epsilon$ and $\delta_z$. Given $\epsilon$ is multiplicative, it corresponds to rescaling the pulse envelope. On the other hand, since $\delta_z$ is additive, the percentage needs a reference point. To standardize the detuning across all qubits, a field strength of 0.0171~GHz, which is typical of IBM $X$ gates, is chosen; this puts the magnitude of $\delta_z$ at $[-3.42, -2.57, -1.71, -8.55, 0, 8.55, 1.71, 2.57, 3.42]\MHz$. Results from QPT are shown in Fig.~\ref{fig::hada}(c-e) and discussed in more detail below.

\subsubsection{Hadamard QPT Results and Analysis}
\label{sec::Hadamard::analysis}
We first address the space curve in \ref{fig::hada}(a); it exhibits two important geometric properties that map it to a \textit{doubly robust} gate. First, it satisfies upfront the closed-curve condition which translates to first-order robustness to detuning errors. Second, this space curve was also further optimized to satisfy the tangent zero-area condition, which translates to first-order robustness to amplitude error as explained in Appendix \ref{app::theory_BARQ}.

The final Hadamard pulse designed by BARQ is shown in Fig.~\ref{fig::hada}(b). This pulse is then scaled to qubit$[16]$ on the \textit{ibm\_strasbourg} backend, meaning a suitable $T_g$ and $\Omega_{\text{Max}}$ are chosen as per Section~\ref{sec::tailoring}. This qubit choice was motivated by the findings in Section~\ref{sec:utility_scale}, as this qubit's leakage insensitivity allows us to shorten the gate times considerably. For this qubit, the gate time is $116\ns$. Further discussion regarding the nuance involved in choosing the gate times can be found in Section~\ref{sec:utility_scale}. The pulse can also be seen to be smooth, with a gradual rise time, and it is slightly off resonance with a detuning of $T_g\Delta = -0.16973$ in the absence of noise.

The experimentally obtained gate infidelities shown in Figs.~\ref{fig::hada}(c,d) were obtained using the methods discussed in Section~\ref{sec::Hadamard::setup}. The SCQC channel fidelity results, shown with orange crosses, can be seen to outperform the IBM gate results, especially as the strength of the induced error increases.

Finally, we reconstruct the Choi matrices at the $[0,10,20]\%$ marks in both $\delta_z$ and $\epsilon$. By illustrating the full channel Choi matrices, we are able to see that the detuning noise, as expected, changes the phases of the Choi matrix entries away from the noise-free case, while the amplitude noise leads to some population shifts as shown by the uneven height in the choi matrix elements.

\subsection{Experimental benchmarking: Single qubit Clifford generation from doubly-robust $X$, $\sqrt{X}$ gates}
\label{sec::X_SX}

\begin{figure*}
    \subfloat{
        \includegraphics[width=0.95\textwidth]{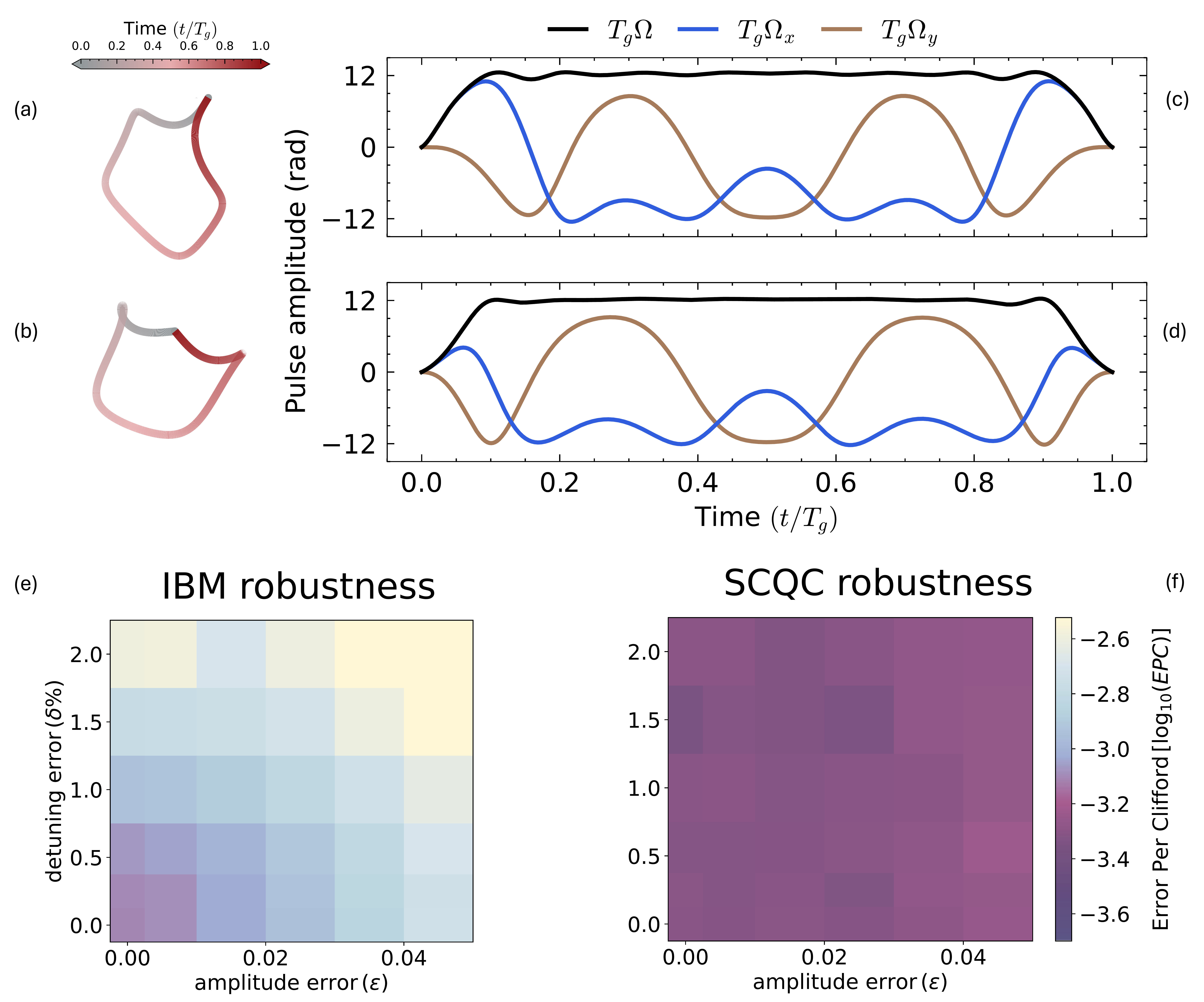}
    }
    \hfill

    \caption{ (a,b) Depictions of the space curves generated by BARQ for doubly robust $X$ and $\sqrt{X}$ gates. Both space curves satisfy the \textit{closed-curve} and \textit{zero-area} conditions. (c,d) The pulse shapes extracted from the space curves before scaling to a given qubit.
    (e,f) Experimental results from qubit[27] of \textit{ibm\_brisbane} with the $(\epsilon,\delta_z)$ EPC heatmaps from standard RB using the standard IBM versus SCQC $X$ and $\sqrt{X}$ transpilation, respectively, of the single-qubit Clifford sequences.}
    \label{fig::heatmap_single}
\end{figure*}

\subsubsection{$X$ and $\sqrt{X}$ RB Setup and Motivation}
\label{sec::X_SX::setup}
For our second demonstration, we can benchmark two gates effectively in a single protocol if we opt for standard randomized benchmarking (RB), making this comparable to the way IBM backends report their gate errors. In standard RB, sequences of Clifford gates are generated and later inverted to leave an overall identity operation. For our SCQC gates, these Clifford sequences reach depths of up to $4000$ Clifford gates for the initial layer, even before the inversion part of the circuit is reached. Once the IBM backend is provided with a Clifford gate, it transpiles it into its standard IBM calibrated gates. In IBM's case, the only physical pulses needed for single-qubit gates generate $\sqrt{X}$ and $X$ gates, while all other single-qubit gates can be obtained by combining these with virtual $Z$ gates that can be applied at no cost~\cite{McKay_2017}. Therefore, to run the Clifford sequences for RB with the SCQC gates, we only need to replace the $\sqrt{X}$ and $ X$ pulses in the IBM transpilations with equivalent SCQC pulses. After this replacement, the error per Clifford (EPC) value we get will be an average of the error per single-qubit Cliffords achievable in practice through SCQC, which we can then compare to the EPC for the IBM scheme.

The narrow interval we utilize for RB involves the sets $[0, 0.5, 1.5, 2.5, 3.5, 4.5]\%$ for multiplicative amplitude error and $[0, 0.25, 0.5, 1.0, 1.5, 2]\%$ for detuning error, which puts the magnitude of the detuning errors at $[0 , 43, 86, 171, 257, 342]\text{ kHz}$. This interval is motivated in Section~\ref{sec::pulse_generation::noise_intervals}. To elaborate further on the parameter calibration uncertainty region, the statistical uncertainty in the frequency calibration is typically on the order of 10s of kHz for IBM's \textit{fine} frequency calibration protocol, and 100s of kHz for their \textit{rough} calibration protocol~\cite{Kanazawa2023_Qiskit_exp}.

To generate robustness heatmaps, we sweep through all the different combinations of $\delta_z$ and $\epsilon$ in the error sets, giving us a heatmap of $36$ unique $(\epsilon,\delta_z)$ combinations. The resulting error per Clifford range of interest, given this narrow-ranged heatmap, is around $7\times10^{-5}$ to $3\times10^{-3}$, which is the practical range over which one typically observes superconducting single-qubit gate fidelities.

Finally, we perform the RB analysis on one noise quadrant, so only considering positive parameter combinations. This is justified by the EPC behavior being  approximately symmetric, as one can also infer from the wider QPT noise sweeps. Second, using one quadrant allows us to obtain a better noise resolution for the heatmap at a reasonable quantum overhead. The results for the heatmaps are shown in Fig.~\ref{fig::heatmap_single} and discussed in more detail below.

\subsubsection{$X$ and $\sqrt{X}$ RB Results and Analysis}
\label{sec::X_SX::analysis}
Following the same protocol used for the Hadamard, BARQ is used to generate the $X$ and $\sqrt{X}$ space curves shown in Figs.~\ref{fig::heatmap_single}(a) and \ref{fig::heatmap_single}(b), respectively. The control pulses obtained from these space curves are shown in Figs.~\ref{fig::heatmap_single}(c) and \ref{fig::heatmap_single}(d) respectively. As with the Hadamard analysis in Section~\ref{sec::Hadamard::analysis} the control pulses are designed to be smooth, short under a reasonable finite drive strength, with practical rise times, and doubly robust. The pulses are also optimized to run nearly on resonance, where the optimal detuning values are $T_g\Delta$ is $(-0.00938, 0.01091)$, respectively.

For the RB experiments, the $X$ and $\sqrt{X}$ pulses are scaled to qubit$[27]$ on \textit{ibm\_brisbane} by choosing an appropriate gate time and driving strength, as we now explain. To demonstrate that it is possible to predict when SCQC pulses are a good alternative to the standard IBM pulses, we pick, for our RB heatmap experiments, a single qubit out of the $127$ qubits available on a single Eagle processor, which was in turn chosen out of the six $127$-qubit devices that were available at the time. This qubit was selected in part because it was known to be relatively less susceptible to leakage, which we inferred from the nearly zero calibrated Derivative Removal by Adiabatic Gate (DRAG) \cite{IBM_drag_2009,IBM_drag_2011, Alexander_2020_Qsikit_pulse} parameter  $(0.0085)$ in the IBM pulse. DRAG pulses are designed specifically to suppress leakage out of the qubit subspace, and the magnitude of their suppression is proportional to this DRAG parameter. Given that the SCQC pulses have not yet been modified to incorporate leakage robustness, this qubit is a good candidate for a direct robustness comparison between the IBM and the SCQC pulses. The relatively low leakage susceptibility on the qubit allows us to run the gates with pulse amplitudes up to $1.35$ times as strong as the IBM $X$ gate's maximum drive amplitude, without the risk of introducing too much error from leakage. This specific gate time is addressed later in this section. We observed a consistent improvement in this reduced leakage susceptibility with each new IBM processor release; if this trend continues, this favorable feature can potentially be even more common on the newer superconducting processors. However, pulse access is not yet offered on these devices.

After scaling the drive strength, the gate times for the $X$ and $\sqrt{X}$ gates are 84.0 ns and 80.0 ns, respectively, which can be compared to the 60 ns durations of the corresponding IBM gates. An initial exploratory gate-time sweep was utilized to help choose the pulse scaling. The purpose of the sweep is to scout the region within which the trade-off is minimal between shortening the gate time and raising the EPC due to the increase in leakage. A more detailed discussion on one way this can be done more systematically is provided in Section~\ref{sec:utility_scale}.

We now examine the heatmaps shown in Fig.~\ref{fig::heatmap_single}(e,f), which depict the EPC at the different noise-level combinations $\epsilon,\delta_z$. This is an example of a case in which the transpilation of the Cliffords under the $X$ and $\sqrt{X}$ SCQC pulses outperforms the IBM built-in pulses across all performance metrics. First, the SCQC EPC remains unchanged at around $4.96\times10^{-4}$ ($\pm5.36\times10^{-5}$) for all the noise-level combinations, barely falling to $5.35\times10^{-4}$ ($\pm7.23\times10^{-5}$) in the largest noise region, which basically still lies within the uncertainty region of the former EPC. Second, not only does the SCQC EPC stay invariant under the noise sweep, but it is also outperforming the IBM EPC in the noise-free case, which corresponds to the origin in the heatmap. The origin is where the IBM gate exhibits its lowest EPC at just $7.82\times10^{-4}$ ($\pm2.40\times10^{-5}$), however it decays to around $3.75\times10^{-3}$ ($\pm6.63\times10^{-4}$) in the noise region $(4\%,2\%)$, the highest $(\epsilon,\delta_z)$ in both noise directions.

It is important to emphasize that, beyond this initial rough sweep, the SCQC gates are used straight from the simulator, meaning the SCQC gates never go through a calibration process. This calibration-free quality is possible because of how insensitive the pulses are to control and device parameter shifts away from their ideal values. Accordingly, for a simulation that is faithful enough to the experiment, if the parameters derived from such a simulation fall within this insensitive region around the ideal gate parameters, the gate would require little to no calibration. This ability to bypass the frequent, time-consuming, and resource-heavy calibration cycles is one of the main practical advantages of our approach. Furthermore, we expect the SCQC pulses to need no further calibration once the gate time is chosen, since the leakage susceptibility of the qubit will depend on the anharmonicity and maximum driving field strength on the qubit.

Finally, we provide several explanations for why the SCQC gate performs better here in the noiseless run as well. The relatively poor performance of the IBM gates on this qubit may be attributed to their poor tuning relative to the real-time device or control parameters at the time of the experiment. The mismatch could have been due to multiple different reasons. Possibilities include a faulty inaccurate calibration process of the gate control parameters, an accurate (but imprecise) calibration with too high an uncertainty in the parameter fitting and measurements used in the calibration process, or an accurate and precise calibration after which the device or control parameters have drifted too much by the time the calibrated gate is used. The key takeaway is that, in all three cases, the SCQC gate is robust to such deviations, hence the favorable performance.

\begin{figure*}
    \subfloat{
        \includegraphics[width=0.95\textwidth]{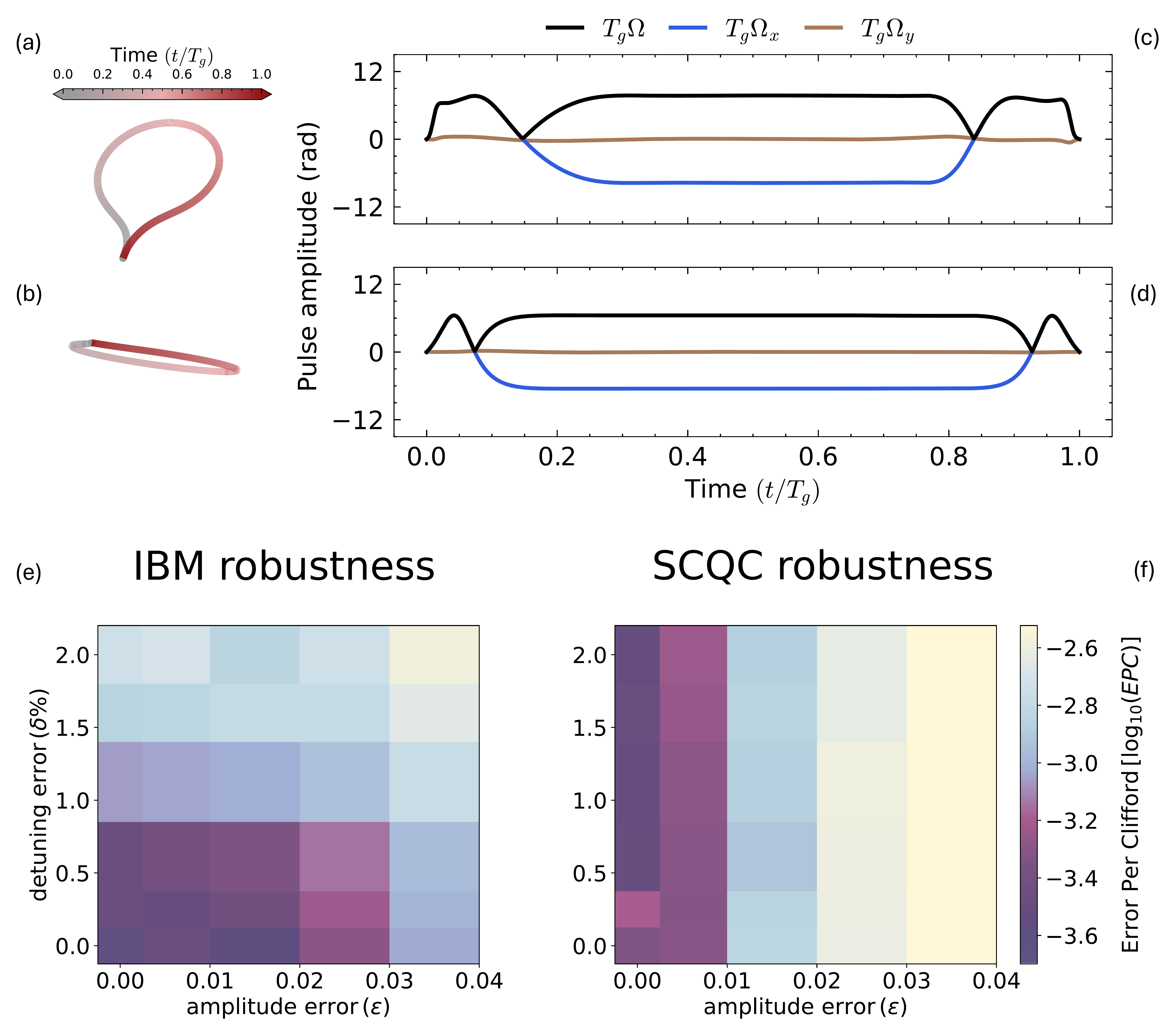}
    }
    \hfill

    \caption{ (a,b) Depictions of the space curves generated by BARQ for the $X$, $\sqrt{X}$ gates. Unlike in Fig.~\ref{fig::heatmap_single}, the space curves here \textbf{only} satisfy the \textit{closed-curve} condition. (c,d) The pulse shapes extracted from the space curve before scaling to a particular qubit.
    (e,f) Experimental results on qubit[0] of \textit{ibm\_brisbane} with the $(\epsilon,\delta_z)$ EPC heatmaps from standard RB using the standard IBM and SCQC $X$, $\sqrt{X}$ transpilation, respectively, of the single-qubit Clifford sequences. The singly robust pulse retains its detuning-robustness along the vertical axis while it loses most of its amplitude-robustness along the horizontal axis.}
    \label{fig::singly_robust}
\end{figure*}

\subsection{Special SCQC use-case: Singly robust gates in exchange for shorter $X$, $\sqrt{X}$ pulses}
\label{sec::single_robust}

\subsubsection{Singly Robust Gates Setup and Motivation}

As discussed earlier in Section \ref{sec::tailoring}, the user is free to allocate weights towards the pulse or gate properties they deem important. In this section, we demonstrate how this can work in practice by following a protocol similar to what is used in Section~\ref{sec::X_SX} for SCQC $X$, $\sqrt{X}$ pulses, only this time we will give very little weight to amplitude robustness, and much more weight to the maximum driving strength (which directly translates to shorter gate times when rescaling for a given maximum driving strength on a particular qubit).

This demonstration serves two purposes. First, we show experimentally how we lose the robustness of the control pulse once its corresponding space curve's geometric robustness condition is loosened. In this case, BARQ does not optimize for the fulfillment of the tangent zero-area condition introduced in Section~\ref{sec::pulse_generation::summary_BARQ_SCQC}, which is the condition that SCQC uses to impose first-order robustness to multiplicative amplitude errors. Second, we again demonstrate the practical versatility of the SCQC formalism and the BARQ control method, since if a user has access to a device where the fluctuations in the driving amplitude fall below a certain practical threshold at which the fidelity is practically unchanged, they can exchange the amplitude robustness for a much shorter pulse. In this case, these singly robust pulses are half as long as the doubly robust $X$, $\sqrt{X}$ pulses discussed in Section~\ref{sec::X_SX} for a given maximum driving field. The results for these singly robust pulses are shown in Fig.~\ref{fig::singly_robust} and discussed in more detail below.

\subsubsection{Singly Robust Gates Results and Analysis}

In Fig.~\ref{fig::singly_robust}(a,b) the space curves that generate the singly robust pulses can be seen to be closed, and so fulfill the condition for first-order robustness to detuning errors. The space curves are also nearly planar and thus lead to drives that are approximately along a single quadrature, as can be see in Fig.~\ref{fig::singly_robust}(c,d). The space curves are also nearly circular, meaning the curvature is nearly constant, which explains why the corresponding control pulses in Fig.~\ref{fig::singly_robust}(c,d) are approaching square-like pulses. This can be understood from Eq.~\ref{eqn:curvature_to_Omega} in Appendix~\ref{app::theory_BARQ}, which gives the correspondence between the driving field and the curvature of the space curve. This matches what we expect from the SCQC formalism, as all these space curve qualities are necessary if we want the shortest possible control pulse under some finite drive~\cite{Zeng_PRA2018}. This use case also touches on the question of how fast we can drive $X$ rotations while still remaining detuning robust. Finally, the EPC heatmaps in Fig.~\ref{fig::singly_robust}(e,f) clearly show how the EPC falls quickly for the SCQC singly-robust gate along the $\epsilon$ axis, a quality we add to the pulse by design. However, the SCQC pulses are still robust to detuning compared to the standard IBM pulses. An important feature here is that the scaling is done on qubit$[0]$ on \textit{ibm\_brisbane} and specifically targets the $60\ns$ gate time. This qubit, like the qubit used in Section~\ref{sec::single_robust}, showed a very low $(0.00973)$ DRAG parameter for its IBM gate calibrations, implying that this qubit's leakage susceptibility was low enough to allow us to drive it at around the same driving strength as the standard IBM $X$ pulse, which was around $0.017834\GHz$, without having to worry about leakage. This again supports our claim that we can predict the qubits on which the SCQC pulses perform well. That said, the SCQC gate's EPC in the noiseless run was around $4.596\times10^{-4}$ ($\pm3.016\times10^{-5}$), while the IBM gate was at $2.604\times10^{-4}$ ($\pm6.904\times10^{-6}$).

\section{Utility scale experiments}
\label{sec:utility_scale}

\subsection{18 Qubit setup}

\begin{figure}\includegraphics[width=0.8\columnwidth]{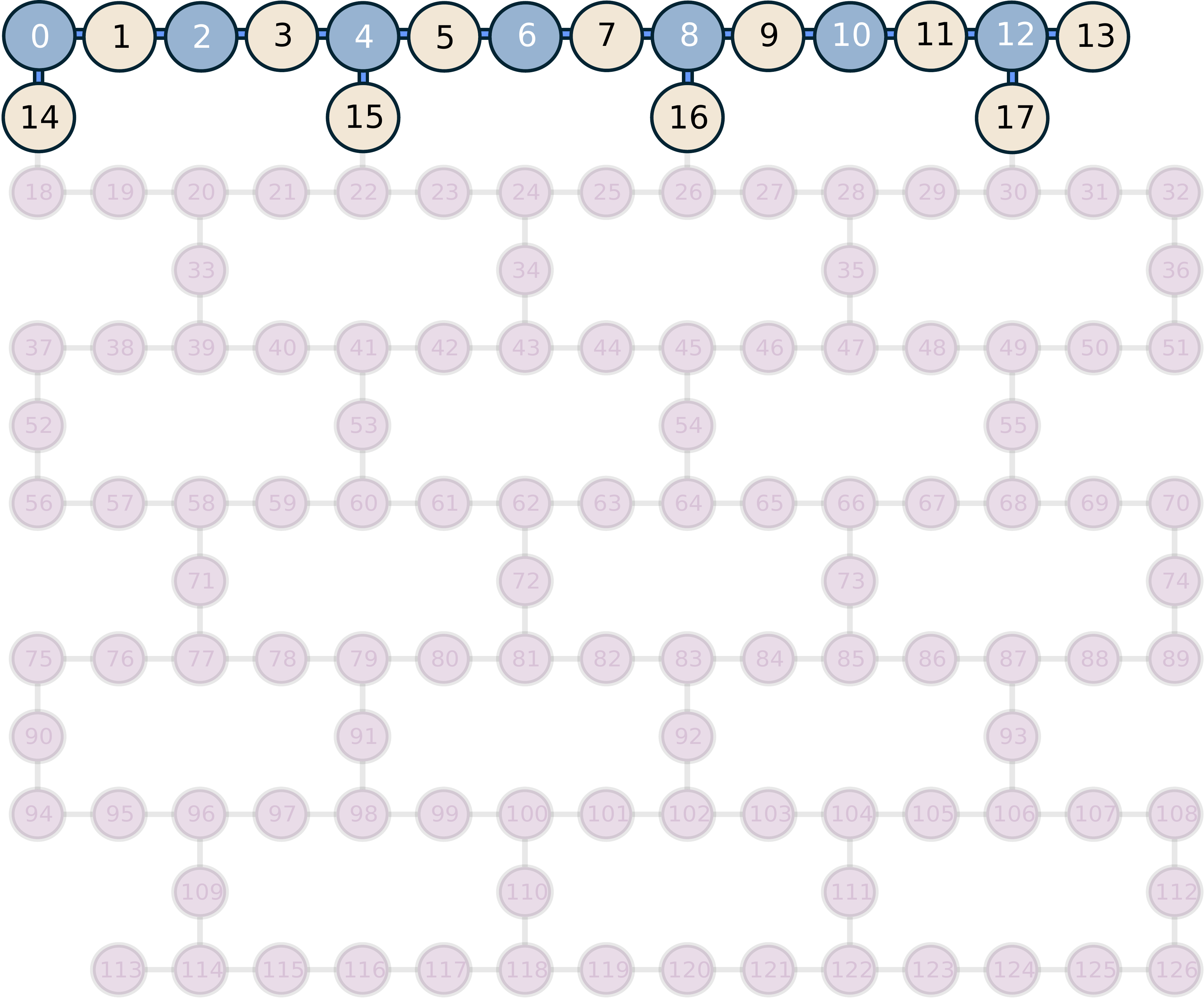}
    \caption{\label{fig:couplingmap}Coupling map for the IBM Eagle processors, including \textit{ibm\_brisbane} and \textit{ibm\_strasbourg} used in this work. The top two rows are highlighted as these rows are the ones used in the utility scale work in Section~\ref{sec:utility_scale}, covering 18 qubits from qubit$[0]$ to qubit$[17]$. Qubits with the same color are pulsed in parallel.}
\end{figure}

To show that we can perform this SCQC control scheme at the utility scale, we repeat the approach presented in Section~\ref{sec::X_SX} in an automated fashion on a subset of $18$ qubits on a different backend. For this demonstration, we used \textit{ibm\_strasbourg} to emphasize that this SCQC behavior was not device specific. To further ensure that the qubit choices are arbitrary, we simply pick the first two rows as highlighted in Fig.~\ref{fig:couplingmap}. By benchmarking pulses on 18 qubits, each of which has its own distinct characteristics, including anharmonicities and leakage susceptibility, we get a good representation of how often the SCQC gates can, in practice, outperform the standard IBM gates and by what margin.

The qubits are divided into the two color groups shown in Fig.~\ref{fig:couplingmap}, such that within each group the qubits are not directly coupled. The following 18-qubit experiments are carried out in two large batches, one for each group. Within each group all experiments are run in parallel, ensuring that the pulses act simultaneously but only on uncoupled qubits. This setup should mimic the state of the newer Heron devices~\cite{IBMQuantum} where the tunable couplers are always off until two-qubit gates are needed, a practice which limits cross-talk. The final pulse scaling for each qubit is then considered on an individual basis, following a gate-time sweep aimed at identifying the gate time with the lowest EPC, which typically lies at the time when the optimum trade-off is seen between more leakage from shorter gate times versus more decoherence from longer gate times. For reference, the ordering of the qubits in terms of gate time is [16, 4, 5, 13, 12, 2, 15, 10, 3, 8, 1, 9, 6, 14, 0, 11, 7,  17] and the corresponding gate times in nanoseconds are [ 88, 112, 132, 144, 156, 168, 188, 188, 188, 192, 192, 196, 212, 212, 212, 220, 220, 220]. It is important to emphasize that all the qubits can accommodate faster gate times, but gate times were chosen specifically to get the lowest possible EPC, whilst not exceeding 220 ns, as we tried to push the EPC to the coherence limit on every qubit up until this 220 ns cutoff. More favorable gate times could have been chosen at the expense of slightly higher gate errors.

\begin{figure*}
        \includegraphics[width=1\textwidth]{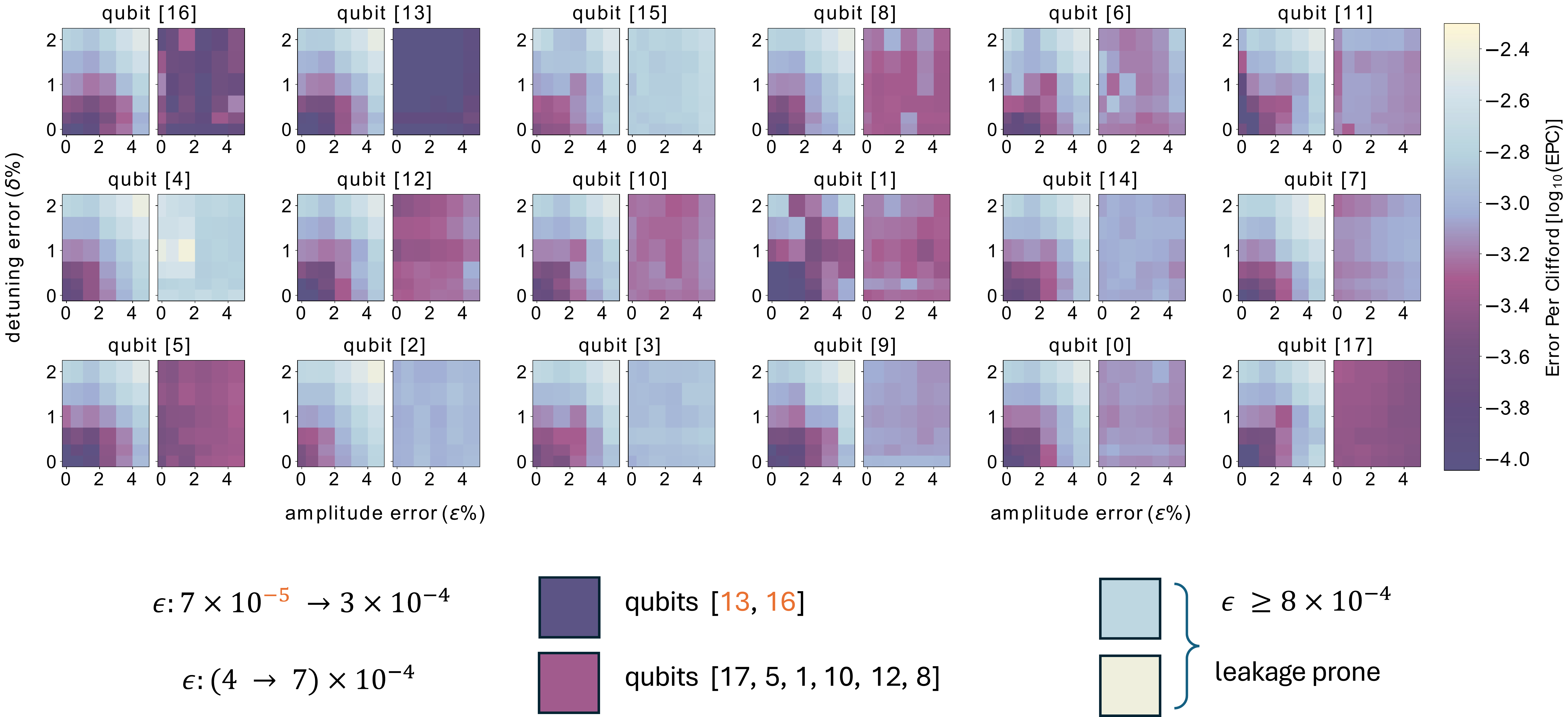}
    \hfill

    \caption{Comparing the error per Clifford obtained from RB for SCQC pulses (right panel in each pair of heatmaps) versus standard IBM pulses (left panel in each pair) as a function of detuning error (vertical axes) and amplitude error (horizontal axes) across the first 18 qubits of \textit{ibm\_strasbourg}. The SCQC pulses outperform IBM pulses for qubits 1,5,8,10,12,17 and especially for qubits 13 and 16. The SCQC pulses under-perform compared to IBM pulses for the remaining qubits due to excessive leakage errors that arise for these qubits.}
    \label{fig::heatmap_utility}
\end{figure*}

\subsection{18 Qubit Results and Analysis}
\label{sec:utility_scale:18_qubit}

Once the gate time, and the corresponding maximum drive strength, are chosen for each of the qubits, the same robustness analysis as in Section~\ref{sec::X_SX} was carried out on the 18 qubits. Fig.~\ref{fig::heatmap_utility} shows these results with 18 pairs of IBM versus SCQC heatmaps, where each pair resembles the heatmaps in Fig.~\ref{fig::heatmap_single}(e,f).

The performance of the SCQC pulses is determined from the EPC obtained while using SCQC $X$, $\sqrt{X}$ pulses to transpile the Clifford gates in the RB sequences. Based on this criterion, the performance of the SCQC pulses on the 18 qubits is grouped into three main categories. First, the indigo region contains two qubits on which SCQC pulses considerably surpassed the IBM gates in terms of robustness whilst still measuring up to the IBM EPC in the noiseless case. That performance was seen on qubits 13 and 16.
For qubit[16], the EPC is around $2\times10^{-4}$, at zero noise and barely changes at all the other noise levels, all while only taking 88 ns (1.47x longer than the IBM pulse, which has a gate time of 60 ns). Qubit[13], although its gate is considerably longer at 144 ns (2.4x of the IBM pulse ), attained the best EPC out of any experiment, measured at $7\times10^{-5}$ for the entirety of the noise region, possibly one of the lowest recorded error rates for DCGs to the best of our knowledge.

The second group is the purple region, which includes qubits 5,12,10,8,1,17, on which the SCQC Clifford transpilation performs well and is considerably more robust than the IBM pulses. However the performance on these qubits in the noiseless region was within $4\times10^{-4}$ to $7\times10^{-4}$, which is still good, but slightly less than the IBM EPC's, which range from $3\times10^{-4}$ to $5\times10^{-4}$.

Finally, we consider the blue and white regions, which are regions where the SCQC gates did not perform as well. These are qubits for which we expected SCQC to under-perform given that their qubit DRAG parameter calibrations were relatively high, implying their larger susceptibility to leakage compared to other qubits, which are the type of qubits our gates are not yet suitable for.

\section{Conclusions and outlook}
\label{sec:conclusions}

In this paper, we experimentally demonstrated the practical utility of Space Curve Quantum Control (SCQC), a method that facilitates the automated design of experimentally practical pulses for fast and smooth dynamically corrected gates (DCGs) robust to multiple noise sources. We used this approach to design Hadamard, $X$, and $\sqrt{X}$ gates that are robust to both detuning noise and pulse-amplitude noise. In order to properly assess the performance of these gates and compare them against standard IBM gates, we implemented benchmarking methods that target the commonly neglected sub-100 kHz frequency noise regime in which common superconducting noise manifests, including the inherent statistical uncertainty in the calibration process. We also used quantum process tomography to show that our SCQC-designed Hadamard gate exhibits superior robustness compared to the standard IBM gate in the presence of high levels of detuning noise. These noise ranges cover the noise bursts in the few MHz range observed from radiation in recent QEC experiments~\cite{acharya2024_QEC_google_subthreshold}. 

The key result in this work, however, is the superior performance of our SCQC-designed gates versus standard IBM gates we observed in randomized benchmarking experiments conducted on 18 qubits of an IBM device, \textit{ibm\_strasbourg}. We observed an error per Clifford (EPC) as low as $7\times10^{-5}$ that persisted across across a broad range of noise strengths up to 342 kHz detuning noise and 4\% amplitude noise acting simultaneously on a qubit. The IBM pulses pale in comparison as their EPC deteriorates down to the order of $3\times10^{-3}$ across the same noise span.. 

Finally, we established that leakage is the limiting factor for these specific SCQC gates, which opens up room for follow-up work in which geometric leakage-robustness conditions could be established within the SCQC framework. The most promising prospect, however, is that we experimentally demonstrated that some qubits are already leakage insensitive enough to allow us to consistently outperform the standard IBM pulses with comparable gate times. Consequently, currently available improvements to qubit leakage sensitivity could turn SCQC's best case performance into the median performance on these large-scale devices, which would give users a utility-scale calibration-free control scheme that is also less sensitive to qubit decoherence.


\begin{acknowledgements}

SEE acknowledges support from the Department of Energy (Grant No. DE-SC0024488).
EB acknowledges support from the Department of Energy (Grant No. DE-SC0022389) and from the Office of Naval Research (Grant No.
N00014-25-1-2125).  This research used resources of the
Oak Ridge Leadership Computing Facility, which is a DOE Office of Science User Facility supported under Contract DE-AC05-00OR22725.

\end{acknowledgements}


\bibliography{SCQC_IBM}

\begin{thebibliography}{64}%
\makeatletter
\providecommand \@ifxundefined [1]{%
 \@ifx{#1\undefined}
}%
\providecommand \@ifnum [1]{%
 \ifnum #1\expandafter \@firstoftwo
 \else \expandafter \@secondoftwo
 \fi
}%
\providecommand \@ifx [1]{%
 \ifx #1\expandafter \@firstoftwo
 \else \expandafter \@secondoftwo
 \fi
}%
\providecommand \natexlab [1]{#1}%
\providecommand \enquote  [1]{``#1''}%
\providecommand \bibnamefont  [1]{#1}%
\providecommand \bibfnamefont [1]{#1}%
\providecommand \citenamefont [1]{#1}%
\providecommand \href@noop [0]{\@secondoftwo}%
\providecommand \href [0]{\begingroup \@sanitize@url \@href}%
\providecommand \@href[1]{\@@startlink{#1}\@@href}%
\providecommand \@@href[1]{\endgroup#1\@@endlink}%
\providecommand \@sanitize@url [0]{\catcode `\\12\catcode `\$12\catcode
  `\&12\catcode `\#12\catcode `\^12\catcode `\_12\catcode `\%12\relax}%
\providecommand \@@startlink[1]{}%
\providecommand \@@endlink[0]{}%
\providecommand \url  [0]{\begingroup\@sanitize@url \@url }%
\providecommand \@url [1]{\endgroup\@href {#1}{\urlprefix }}%
\providecommand \urlprefix  [0]{URL }%
\providecommand \Eprint [0]{\href }%
\providecommand \doibase [0]{https://doi.org/}%
\providecommand \selectlanguage [0]{\@gobble}%
\providecommand \bibinfo  [0]{\@secondoftwo}%
\providecommand \bibfield  [0]{\@secondoftwo}%
\providecommand \translation [1]{[#1]}%
\providecommand \BibitemOpen [0]{}%
\providecommand \bibitemStop [0]{}%
\providecommand \bibitemNoStop [0]{.\EOS\space}%
\providecommand \EOS [0]{\spacefactor3000\relax}%
\providecommand \BibitemShut  [1]{\csname bibitem#1\endcsname}%
\let\auto@bib@innerbib\@empty
\bibitem [{\citenamefont {Acharya}\ \emph {et~al.}(2024)\citenamefont
  {Acharya}, \citenamefont {Abanin}, \citenamefont {Aghababaie-Beni},
  \citenamefont {Aleiner}, \citenamefont {Andersen}, \citenamefont {Ansmann},
  \citenamefont {Arute}, \citenamefont {Arya}, \citenamefont {Asfaw},
  \citenamefont {Astrakhantsev} \emph
  {et~al.}}]{acharya2024_QEC_google_subthreshold}%
  \BibitemOpen
  \bibfield  {author} {\bibinfo {author} {\bibfnamefont {R.}~\bibnamefont
  {Acharya}}, \bibinfo {author} {\bibfnamefont {D.~A.}\ \bibnamefont {Abanin}},
  \bibinfo {author} {\bibfnamefont {L.}~\bibnamefont {Aghababaie-Beni}},
  \bibinfo {author} {\bibfnamefont {I.}~\bibnamefont {Aleiner}}, \bibinfo
  {author} {\bibfnamefont {T.~I.}\ \bibnamefont {Andersen}}, \bibinfo {author}
  {\bibfnamefont {M.}~\bibnamefont {Ansmann}}, \bibinfo {author} {\bibfnamefont
  {F.}~\bibnamefont {Arute}}, \bibinfo {author} {\bibfnamefont
  {K.}~\bibnamefont {Arya}}, \bibinfo {author} {\bibfnamefont {A.}~\bibnamefont
  {Asfaw}}, \bibinfo {author} {\bibnamefont {Astrakhantsev}}, \emph {et~al.},\
  }\bibfield  {title} {\bibinfo {title} {Quantum error correction below the
  surface code threshold},\ }\href {https://doi.org/10.1038/s41586-024-08449-y}
  {\bibfield  {journal} {\bibinfo  {journal} {Nature}\ }\textbf {\bibinfo
  {volume} {638}},\ \bibinfo {pages} {920–926} (\bibinfo {year}
  {2024})}\BibitemShut {NoStop}%
\bibitem [{\citenamefont {Kim}\ \emph {et~al.}(2023)\citenamefont {Kim},
  \citenamefont {Eddins}, \citenamefont {Anand}, \citenamefont {Wei},
  \citenamefont {van~den Berg}, \citenamefont {Rosenblatt}, \citenamefont
  {Nayfeh}, \citenamefont {Wu}, \citenamefont {Zaletel}, \citenamefont
  {Temme},\ and\ \citenamefont {Kandala}}]{Kim2023}%
  \BibitemOpen
  \bibfield  {author} {\bibinfo {author} {\bibfnamefont {Y.}~\bibnamefont
  {Kim}}, \bibinfo {author} {\bibfnamefont {A.}~\bibnamefont {Eddins}},
  \bibinfo {author} {\bibfnamefont {S.}~\bibnamefont {Anand}}, \bibinfo
  {author} {\bibfnamefont {K.~X.}\ \bibnamefont {Wei}}, \bibinfo {author}
  {\bibfnamefont {E.}~\bibnamefont {van~den Berg}}, \bibinfo {author}
  {\bibfnamefont {S.}~\bibnamefont {Rosenblatt}}, \bibinfo {author}
  {\bibfnamefont {H.}~\bibnamefont {Nayfeh}}, \bibinfo {author} {\bibfnamefont
  {Y.}~\bibnamefont {Wu}}, \bibinfo {author} {\bibfnamefont {M.}~\bibnamefont
  {Zaletel}}, \bibinfo {author} {\bibfnamefont {K.}~\bibnamefont {Temme}},\
  and\ \bibinfo {author} {\bibfnamefont {A.}~\bibnamefont {Kandala}},\
  }\bibfield  {title} {\bibinfo {title} {Evidence for the utility of quantum
  computing before fault tolerance},\ }\bibfield  {journal} {\bibinfo
  {journal} {Nature}\ }\textbf {\bibinfo {volume} {618}},\ \href
  {https://doi.org/10.1038/s41586-023-06096-3} {10.1038/s41586-023-06096-3}
  (\bibinfo {year} {2023})\BibitemShut {NoStop}%
\bibitem [{IBM()}]{IBMQuantum}%
  \BibitemOpen
  \href@noop {} {\bibinfo {title} {Ibm quantum}},\ \bibinfo {howpublished}
  {\url{ https://quantum.ibm.com/}},\ \bibinfo {note} {accessed:
  2025}\BibitemShut {NoStop}%
\bibitem [{\citenamefont {Tripathi}\ \emph {et~al.}(2024)\citenamefont
  {Tripathi}, \citenamefont {Chen}, \citenamefont {Levenson-Falk},\ and\
  \citenamefont {Lidar}}]{Tripathi_2024}%
  \BibitemOpen
  \bibfield  {author} {\bibinfo {author} {\bibfnamefont {V.}~\bibnamefont
  {Tripathi}}, \bibinfo {author} {\bibfnamefont {H.}~\bibnamefont {Chen}},
  \bibinfo {author} {\bibfnamefont {E.}~\bibnamefont {Levenson-Falk}},\ and\
  \bibinfo {author} {\bibfnamefont {D.~A.}\ \bibnamefont {Lidar}},\ }\bibfield
  {title} {\bibinfo {title} {Modeling low- and high-frequency noise in transmon
  qubits with resource-efficient measurement},\ }\bibfield  {journal} {\bibinfo
   {journal} {PRX Quantum}\ }\textbf {\bibinfo {volume} {5}},\ \href
  {https://doi.org/10.1103/prxquantum.5.010320} {10.1103/prxquantum.5.010320}
  (\bibinfo {year} {2024})\BibitemShut {NoStop}%
\bibitem [{\citenamefont {Burnett}\ \emph {et~al.}(2019)\citenamefont
  {Burnett}, \citenamefont {Bengtsson}, \citenamefont {Scigliuzzo},
  \citenamefont {Niepce}, \citenamefont {Kudra}, \citenamefont {Delsing},\ and\
  \citenamefont {Bylander}}]{Burnett_2019_study_dephasing_with_DD}%
  \BibitemOpen
  \bibfield  {author} {\bibinfo {author} {\bibfnamefont {J.~J.}\ \bibnamefont
  {Burnett}}, \bibinfo {author} {\bibfnamefont {A.}~\bibnamefont {Bengtsson}},
  \bibinfo {author} {\bibfnamefont {M.}~\bibnamefont {Scigliuzzo}}, \bibinfo
  {author} {\bibfnamefont {D.}~\bibnamefont {Niepce}}, \bibinfo {author}
  {\bibfnamefont {M.}~\bibnamefont {Kudra}}, \bibinfo {author} {\bibfnamefont
  {P.}~\bibnamefont {Delsing}},\ and\ \bibinfo {author} {\bibfnamefont
  {J.}~\bibnamefont {Bylander}},\ }\bibfield  {title} {\bibinfo {title}
  {Decoherence benchmarking of superconducting qubits},\ }\bibfield  {journal}
  {\bibinfo  {journal} {npj Quantum Information}\ }\textbf {\bibinfo {volume}
  {5}},\ \href {https://doi.org/10.1038/s41534-019-0168-5}
  {10.1038/s41534-019-0168-5} (\bibinfo {year} {2019})\BibitemShut {NoStop}%
\bibitem [{\citenamefont {Wellstood}\ \emph {et~al.}(1987)\citenamefont
  {Wellstood}, \citenamefont {Urbina},\ and\ \citenamefont
  {Clarke}}]{1/f_0.58-0.8_1987}%
  \BibitemOpen
  \bibfield  {author} {\bibinfo {author} {\bibfnamefont {F.~C.}\ \bibnamefont
  {Wellstood}}, \bibinfo {author} {\bibfnamefont {C.}~\bibnamefont {Urbina}},\
  and\ \bibinfo {author} {\bibfnamefont {J.}~\bibnamefont {Clarke}},\
  }\bibfield  {title} {\bibinfo {title} {Low‐frequency noise in dc
  superconducting quantum interference devices below 1 k},\ }\href
  {https://doi.org/10.1063/1.98041} {\bibfield  {journal} {\bibinfo  {journal}
  {Applied Physics Letters}\ }\textbf {\bibinfo {volume} {50}},\ \bibinfo
  {pages} {772} (\bibinfo {year} {1987})},\ \Eprint
  {https://arxiv.org/abs/https://pubs.aip.org/aip/apl/article-pdf/50/12/772/18460648/772\_1\_online.pdf}
  {https://pubs.aip.org/aip/apl/article-pdf/50/12/772/18460648/772\_1\_online.pdf}
  \BibitemShut {NoStop}%
\bibitem [{\citenamefont {Rower}\ \emph {et~al.}(2023)\citenamefont {Rower},
  \citenamefont {Ateshian}, \citenamefont {Li}, \citenamefont {Hays},
  \citenamefont {Bluvstein}, \citenamefont {Ding}, \citenamefont {Kannan},
  \citenamefont {Almanakly}, \citenamefont {Braumüller}, \citenamefont {Kim},
  \citenamefont {Melville}, \citenamefont {Niedzielski}, \citenamefont
  {Schwartz}, \citenamefont {Yoder}, \citenamefont {Orlando}, \citenamefont
  {Wang}, \citenamefont {Gustavsson}, \citenamefont {Grover}, \citenamefont
  {Serniak}, \citenamefont {Comin},\ and\ \citenamefont
  {Oliver}}]{Rower_2023_1/f_supression}%
  \BibitemOpen
  \bibfield  {author} {\bibinfo {author} {\bibfnamefont {D.~A.}\ \bibnamefont
  {Rower}}, \bibinfo {author} {\bibfnamefont {L.}~\bibnamefont {Ateshian}},
  \bibinfo {author} {\bibfnamefont {L.~H.}\ \bibnamefont {Li}}, \bibinfo
  {author} {\bibfnamefont {M.}~\bibnamefont {Hays}}, \bibinfo {author}
  {\bibfnamefont {D.}~\bibnamefont {Bluvstein}}, \bibinfo {author}
  {\bibfnamefont {L.}~\bibnamefont {Ding}}, \bibinfo {author} {\bibfnamefont
  {B.}~\bibnamefont {Kannan}}, \bibinfo {author} {\bibfnamefont
  {A.}~\bibnamefont {Almanakly}}, \bibinfo {author} {\bibfnamefont
  {J.}~\bibnamefont {Braumüller}}, \bibinfo {author} {\bibfnamefont {D.~K.}\
  \bibnamefont {Kim}}, \bibinfo {author} {\bibfnamefont {A.}~\bibnamefont
  {Melville}}, \bibinfo {author} {\bibfnamefont {B.~M.}\ \bibnamefont
  {Niedzielski}}, \bibinfo {author} {\bibfnamefont {M.~E.}\ \bibnamefont
  {Schwartz}}, \bibinfo {author} {\bibfnamefont {J.~L.}\ \bibnamefont {Yoder}},
  \bibinfo {author} {\bibfnamefont {T.~P.}\ \bibnamefont {Orlando}}, \bibinfo
  {author} {\bibfnamefont {J.~I.-J.}\ \bibnamefont {Wang}}, \bibinfo {author}
  {\bibfnamefont {S.}~\bibnamefont {Gustavsson}}, \bibinfo {author}
  {\bibfnamefont {J.~A.}\ \bibnamefont {Grover}}, \bibinfo {author}
  {\bibfnamefont {K.}~\bibnamefont {Serniak}}, \bibinfo {author} {\bibfnamefont
  {R.}~\bibnamefont {Comin}},\ and\ \bibinfo {author} {\bibfnamefont {W.~D.}\
  \bibnamefont {Oliver}},\ }\bibfield  {title} {\bibinfo {title} {Evolution of
  1/f flux noise in superconducting qubits with weak magnetic fields},\
  }\bibfield  {journal} {\bibinfo  {journal} {Physical Review Letters}\
  }\textbf {\bibinfo {volume} {130}},\ \href
  {https://doi.org/10.1103/physrevlett.130.220602}
  {10.1103/physrevlett.130.220602} (\bibinfo {year} {2023})\BibitemShut
  {NoStop}%
\bibitem [{\citenamefont {Christensen}\ \emph {et~al.}(2019)\citenamefont
  {Christensen}, \citenamefont {Wilen}, \citenamefont {Opremcak}, \citenamefont
  {Nelson}, \citenamefont {Schlenker}, \citenamefont {Zimonick}, \citenamefont
  {Faoro}, \citenamefont {Ioffe}, \citenamefont {Rosen}, \citenamefont
  {DuBois}, \citenamefont {Plourde},\ and\ \citenamefont
  {McDermott}}]{Christensen_2019_1/f_1.9}%
  \BibitemOpen
  \bibfield  {author} {\bibinfo {author} {\bibfnamefont {B.~G.}\ \bibnamefont
  {Christensen}}, \bibinfo {author} {\bibfnamefont {C.~D.}\ \bibnamefont
  {Wilen}}, \bibinfo {author} {\bibfnamefont {A.}~\bibnamefont {Opremcak}},
  \bibinfo {author} {\bibfnamefont {J.}~\bibnamefont {Nelson}}, \bibinfo
  {author} {\bibfnamefont {F.}~\bibnamefont {Schlenker}}, \bibinfo {author}
  {\bibfnamefont {C.~H.}\ \bibnamefont {Zimonick}}, \bibinfo {author}
  {\bibfnamefont {L.}~\bibnamefont {Faoro}}, \bibinfo {author} {\bibfnamefont
  {L.~B.}\ \bibnamefont {Ioffe}}, \bibinfo {author} {\bibfnamefont {Y.~J.}\
  \bibnamefont {Rosen}}, \bibinfo {author} {\bibfnamefont {J.~L.}\ \bibnamefont
  {DuBois}}, \bibinfo {author} {\bibfnamefont {B.~L.~T.}\ \bibnamefont
  {Plourde}},\ and\ \bibinfo {author} {\bibfnamefont {R.}~\bibnamefont
  {McDermott}},\ }\bibfield  {title} {\bibinfo {title} {Anomalous charge noise
  in superconducting qubits},\ }\bibfield  {journal} {\bibinfo  {journal}
  {Physical Review B}\ }\textbf {\bibinfo {volume} {100}},\ \href
  {https://doi.org/10.1103/physrevb.100.140503} {10.1103/physrevb.100.140503}
  (\bibinfo {year} {2019})\BibitemShut {NoStop}%
\bibitem [{\citenamefont {Paladino}\ \emph {et~al.}(2014)\citenamefont
  {Paladino}, \citenamefont {Galperin}, \citenamefont {Falci},\ and\
  \citenamefont {Altshuler}}]{1/f_review_RevModPhys.86.361}%
  \BibitemOpen
  \bibfield  {author} {\bibinfo {author} {\bibfnamefont {E.}~\bibnamefont
  {Paladino}}, \bibinfo {author} {\bibfnamefont {Y.~M.}\ \bibnamefont
  {Galperin}}, \bibinfo {author} {\bibfnamefont {G.}~\bibnamefont {Falci}},\
  and\ \bibinfo {author} {\bibfnamefont {B.~L.}\ \bibnamefont {Altshuler}},\
  }\bibfield  {title} {\bibinfo {title} {1/f noise:
  Implications for solid-state quantum information},\ }\href
  {https://doi.org/10.1103/RevModPhys.86.361} {\bibfield  {journal} {\bibinfo
  {journal} {Rev. Mod. Phys.}\ }\textbf {\bibinfo {volume} {86}},\ \bibinfo
  {pages} {361} (\bibinfo {year} {2014})}\BibitemShut {NoStop}%
\bibitem [{\citenamefont {Krantz}\ \emph {et~al.}(2019)\citenamefont {Krantz},
  \citenamefont {Kjaergaard}, \citenamefont {Yan}, \citenamefont {Orlando},
  \citenamefont {Gustavsson},\ and\ \citenamefont
  {Oliver}}]{Krantz_2019_engineer_guide}%
  \BibitemOpen
  \bibfield  {author} {\bibinfo {author} {\bibfnamefont {P.}~\bibnamefont
  {Krantz}}, \bibinfo {author} {\bibfnamefont {M.}~\bibnamefont {Kjaergaard}},
  \bibinfo {author} {\bibfnamefont {F.}~\bibnamefont {Yan}}, \bibinfo {author}
  {\bibfnamefont {T.~P.}\ \bibnamefont {Orlando}}, \bibinfo {author}
  {\bibfnamefont {S.}~\bibnamefont {Gustavsson}},\ and\ \bibinfo {author}
  {\bibfnamefont {W.~D.}\ \bibnamefont {Oliver}},\ }\bibfield  {title}
  {\bibinfo {title} {A quantum engineer’s guide to superconducting qubits},\
  }\bibfield  {journal} {\bibinfo  {journal} {Applied Physics Reviews}\
  }\textbf {\bibinfo {volume} {6}},\ \href {https://doi.org/10.1063/1.5089550}
  {10.1063/1.5089550} (\bibinfo {year} {2019})\BibitemShut {NoStop}%
\bibitem [{\citenamefont {Kumar}\ \emph {et~al.}(2016)\citenamefont {Kumar},
  \citenamefont {Sendelbach}, \citenamefont {Beck}, \citenamefont {Freeland},
  \citenamefont {Wang}, \citenamefont {Wang}, \citenamefont {Yu}, \citenamefont
  {Wu}, \citenamefont {Pappas},\ and\ \citenamefont
  {McDermott}}]{Kumar_2016_1/f_flux_origin}%
  \BibitemOpen
  \bibfield  {author} {\bibinfo {author} {\bibfnamefont {P.}~\bibnamefont
  {Kumar}}, \bibinfo {author} {\bibfnamefont {S.}~\bibnamefont {Sendelbach}},
  \bibinfo {author} {\bibfnamefont {M.}~\bibnamefont {Beck}}, \bibinfo {author}
  {\bibfnamefont {J.}~\bibnamefont {Freeland}}, \bibinfo {author}
  {\bibfnamefont {Z.}~\bibnamefont {Wang}}, \bibinfo {author} {\bibfnamefont
  {H.}~\bibnamefont {Wang}}, \bibinfo {author} {\bibfnamefont {C.~C.}\
  \bibnamefont {Yu}}, \bibinfo {author} {\bibfnamefont {R.}~\bibnamefont {Wu}},
  \bibinfo {author} {\bibfnamefont {D.}~\bibnamefont {Pappas}},\ and\ \bibinfo
  {author} {\bibfnamefont {R.}~\bibnamefont {McDermott}},\ }\bibfield  {title}
  {\bibinfo {title} {Origin and reduction of 1/f magnetic flux noise in
  superconducting devices},\ }\bibfield  {journal} {\bibinfo  {journal}
  {Physical Review Applied}\ }\textbf {\bibinfo {volume} {6}},\ \href
  {https://doi.org/10.1103/physrevapplied.6.041001}
  {10.1103/physrevapplied.6.041001} (\bibinfo {year} {2016})\BibitemShut
  {NoStop}%
\bibitem [{\citenamefont {Burnett}\ \emph {et~al.}(2014)\citenamefont
  {Burnett}, \citenamefont {Faoro}, \citenamefont {Wisby}, \citenamefont
  {Gurtovoi}, \citenamefont {Chernykh}, \citenamefont {Mikhailov},
  \citenamefont {Tulin}, \citenamefont {Shaikhaidarov}, \citenamefont
  {Antonov}, \citenamefont {Meeson}, \citenamefont {Tzalenchuk},\ and\
  \citenamefont {Lindström}}]{Burnett_2014}%
  \BibitemOpen
  \bibfield  {author} {\bibinfo {author} {\bibfnamefont {J.}~\bibnamefont
  {Burnett}}, \bibinfo {author} {\bibfnamefont {L.}~\bibnamefont {Faoro}},
  \bibinfo {author} {\bibfnamefont {I.}~\bibnamefont {Wisby}}, \bibinfo
  {author} {\bibfnamefont {V.~L.}\ \bibnamefont {Gurtovoi}}, \bibinfo {author}
  {\bibfnamefont {A.~V.}\ \bibnamefont {Chernykh}}, \bibinfo {author}
  {\bibfnamefont {G.~M.}\ \bibnamefont {Mikhailov}}, \bibinfo {author}
  {\bibfnamefont {V.~A.}\ \bibnamefont {Tulin}}, \bibinfo {author}
  {\bibfnamefont {R.}~\bibnamefont {Shaikhaidarov}}, \bibinfo {author}
  {\bibfnamefont {V.}~\bibnamefont {Antonov}}, \bibinfo {author} {\bibfnamefont
  {P.~J.}\ \bibnamefont {Meeson}}, \bibinfo {author} {\bibfnamefont {A.~Y.}\
  \bibnamefont {Tzalenchuk}},\ and\ \bibinfo {author} {\bibfnamefont
  {T.}~\bibnamefont {Lindström}},\ }\bibfield  {title} {\bibinfo {title}
  {Evidence for interacting two-level systems from the 1/f noise of a
  superconducting resonator},\ }\bibfield  {journal} {\bibinfo  {journal}
  {Nature Communications}\ }\textbf {\bibinfo {volume} {5}},\ \href
  {https://doi.org/10.1038/ncomms5119} {10.1038/ncomms5119} (\bibinfo {year}
  {2014})\BibitemShut {NoStop}%
\bibitem [{\citenamefont {Houck}\ \emph {et~al.}(2009)\citenamefont {Houck},
  \citenamefont {Koch}, \citenamefont {Devoret}, \citenamefont {Girvin},\ and\
  \citenamefont {Schoelkopf}}]{Houck_2009}%
  \BibitemOpen
  \bibfield  {author} {\bibinfo {author} {\bibfnamefont {A.~A.}\ \bibnamefont
  {Houck}}, \bibinfo {author} {\bibfnamefont {J.}~\bibnamefont {Koch}},
  \bibinfo {author} {\bibfnamefont {M.~H.}\ \bibnamefont {Devoret}}, \bibinfo
  {author} {\bibfnamefont {S.~M.}\ \bibnamefont {Girvin}},\ and\ \bibinfo
  {author} {\bibfnamefont {R.~J.}\ \bibnamefont {Schoelkopf}},\ }\bibfield
  {title} {\bibinfo {title} {Life after charge noise: recent results with
  transmon qubits},\ }\href {https://doi.org/10.1007/s11128-009-0100-6}
  {\bibfield  {journal} {\bibinfo  {journal} {Quantum Information Processing}\
  }\textbf {\bibinfo {volume} {8}},\ \bibinfo {pages} {105–115} (\bibinfo
  {year} {2009})}\BibitemShut {NoStop}%
\bibitem [{\citenamefont {Klimov}\ \emph {et~al.}(2018)\citenamefont {Klimov},
  \citenamefont {Kelly}, \citenamefont {Chen}, \citenamefont {Neeley},
  \citenamefont {Megrant}, \citenamefont {Burkett}, \citenamefont {Barends},
  \citenamefont {Arya}, \citenamefont {Chiaro}, \citenamefont {Chen},
  \citenamefont {Dunsworth}, \citenamefont {Fowler}, \citenamefont {Foxen},
  \citenamefont {Gidney}, \citenamefont {Giustina}, \citenamefont {Graff},
  \citenamefont {Huang}, \citenamefont {Jeffrey}, \citenamefont {Lucero},
  \citenamefont {Mutus}, \citenamefont {Naaman}, \citenamefont {Neill},
  \citenamefont {Quintana}, \citenamefont {Roushan}, \citenamefont {Sank},
  \citenamefont {Vainsencher}, \citenamefont {Wenner}, \citenamefont {White},
  \citenamefont {Boixo}, \citenamefont {Babbush}, \citenamefont {Smelyanskiy},
  \citenamefont {Neven},\ and\ \citenamefont
  {Martinis}}]{Klimov_2018_tls_qubit_fluctuations}%
  \BibitemOpen
  \bibfield  {author} {\bibinfo {author} {\bibfnamefont {P.}~\bibnamefont
  {Klimov}}, \bibinfo {author} {\bibfnamefont {J.}~\bibnamefont {Kelly}},
  \bibinfo {author} {\bibfnamefont {Z.}~\bibnamefont {Chen}}, \bibinfo {author}
  {\bibfnamefont {M.}~\bibnamefont {Neeley}}, \bibinfo {author} {\bibfnamefont
  {A.}~\bibnamefont {Megrant}}, \bibinfo {author} {\bibfnamefont
  {B.}~\bibnamefont {Burkett}}, \bibinfo {author} {\bibfnamefont
  {R.}~\bibnamefont {Barends}}, \bibinfo {author} {\bibfnamefont
  {K.}~\bibnamefont {Arya}}, \bibinfo {author} {\bibfnamefont {B.}~\bibnamefont
  {Chiaro}}, \bibinfo {author} {\bibfnamefont {Y.}~\bibnamefont {Chen}},
  \bibinfo {author} {\bibfnamefont {A.}~\bibnamefont {Dunsworth}}, \bibinfo
  {author} {\bibfnamefont {A.}~\bibnamefont {Fowler}}, \bibinfo {author}
  {\bibfnamefont {B.}~\bibnamefont {Foxen}}, \bibinfo {author} {\bibfnamefont
  {C.}~\bibnamefont {Gidney}}, \bibinfo {author} {\bibfnamefont
  {M.}~\bibnamefont {Giustina}}, \bibinfo {author} {\bibfnamefont
  {R.}~\bibnamefont {Graff}}, \bibinfo {author} {\bibfnamefont
  {T.}~\bibnamefont {Huang}}, \bibinfo {author} {\bibfnamefont
  {E.}~\bibnamefont {Jeffrey}}, \bibinfo {author} {\bibfnamefont
  {E.}~\bibnamefont {Lucero}}, \bibinfo {author} {\bibfnamefont
  {J.}~\bibnamefont {Mutus}}, \bibinfo {author} {\bibfnamefont
  {O.}~\bibnamefont {Naaman}}, \bibinfo {author} {\bibfnamefont
  {C.}~\bibnamefont {Neill}}, \bibinfo {author} {\bibfnamefont
  {C.}~\bibnamefont {Quintana}}, \bibinfo {author} {\bibfnamefont
  {P.}~\bibnamefont {Roushan}}, \bibinfo {author} {\bibfnamefont
  {D.}~\bibnamefont {Sank}}, \bibinfo {author} {\bibfnamefont {A.}~\bibnamefont
  {Vainsencher}}, \bibinfo {author} {\bibfnamefont {J.}~\bibnamefont {Wenner}},
  \bibinfo {author} {\bibfnamefont {T.}~\bibnamefont {White}}, \bibinfo
  {author} {\bibfnamefont {S.}~\bibnamefont {Boixo}}, \bibinfo {author}
  {\bibfnamefont {R.}~\bibnamefont {Babbush}}, \bibinfo {author} {\bibfnamefont
  {V.}~\bibnamefont {Smelyanskiy}}, \bibinfo {author} {\bibfnamefont
  {H.}~\bibnamefont {Neven}},\ and\ \bibinfo {author} {\bibfnamefont
  {J.}~\bibnamefont {Martinis}},\ }\bibfield  {title} {\bibinfo {title}
  {Fluctuations of energy-relaxation times in superconducting qubits},\
  }\bibfield  {journal} {\bibinfo  {journal} {Physical Review Letters}\
  }\textbf {\bibinfo {volume} {121}},\ \href
  {https://doi.org/10.1103/physrevlett.121.090502}
  {10.1103/physrevlett.121.090502} (\bibinfo {year} {2018})\BibitemShut
  {NoStop}%
\bibitem [{\citenamefont {Schl\"or}\ \emph {et~al.}(2019)\citenamefont
  {Schl\"or}, \citenamefont {Lisenfeld}, \citenamefont {M\"uller},
  \citenamefont {Bilmes}, \citenamefont {Schneider}, \citenamefont {Pappas},
  \citenamefont {Ustinov},\ and\ \citenamefont
  {Weides}}]{decoherence_IBM_TLS_PRL}%
  \BibitemOpen
  \bibfield  {author} {\bibinfo {author} {\bibfnamefont {S.}~\bibnamefont
  {Schl\"or}}, \bibinfo {author} {\bibfnamefont {J.}~\bibnamefont {Lisenfeld}},
  \bibinfo {author} {\bibfnamefont {C.}~\bibnamefont {M\"uller}}, \bibinfo
  {author} {\bibfnamefont {A.}~\bibnamefont {Bilmes}}, \bibinfo {author}
  {\bibfnamefont {A.}~\bibnamefont {Schneider}}, \bibinfo {author}
  {\bibfnamefont {D.~P.}\ \bibnamefont {Pappas}}, \bibinfo {author}
  {\bibfnamefont {A.~V.}\ \bibnamefont {Ustinov}},\ and\ \bibinfo {author}
  {\bibfnamefont {M.}~\bibnamefont {Weides}},\ }\bibfield  {title} {\bibinfo
  {title} {Correlating decoherence in transmon qubits: Low frequency noise by
  single fluctuators},\ }\href {https://doi.org/10.1103/PhysRevLett.123.190502}
  {\bibfield  {journal} {\bibinfo  {journal} {Phys. Rev. Lett.}\ }\textbf
  {\bibinfo {volume} {123}},\ \bibinfo {pages} {190502} (\bibinfo {year}
  {2019})}\BibitemShut {NoStop}%
\bibitem [{\citenamefont {McEwen}\ \emph {et~al.}(2024)\citenamefont {McEwen},
  \citenamefont {Miao}, \citenamefont {Atalaya}, \citenamefont {Bilmes},
  \citenamefont {Crook}, \citenamefont {Bovaird}, \citenamefont {Kreikebaum},
  \citenamefont {Zobrist}, \citenamefont {Jeffrey}, \citenamefont {Ying},
  \citenamefont {Bengtsson}, \citenamefont {Chang}, \citenamefont {Dunsworth},
  \citenamefont {Kelly}, \citenamefont {Zhang}, \citenamefont {Forati},
  \citenamefont {Acharya}, \citenamefont {Iveland}, \citenamefont {Liu},
  \citenamefont {Kim}, \citenamefont {Burkett}, \citenamefont {Megrant},
  \citenamefont {Chen}, \citenamefont {Neill}, \citenamefont {Sank},
  \citenamefont {Devoret},\ and\ \citenamefont
  {Opremcak}}]{high-energy_impacts_TLS_google}%
  \BibitemOpen
  \bibfield  {author} {\bibinfo {author} {\bibfnamefont {M.}~\bibnamefont
  {McEwen}}, \bibinfo {author} {\bibfnamefont {K.~C.}\ \bibnamefont {Miao}},
  \bibinfo {author} {\bibfnamefont {J.}~\bibnamefont {Atalaya}}, \bibinfo
  {author} {\bibfnamefont {A.}~\bibnamefont {Bilmes}}, \bibinfo {author}
  {\bibfnamefont {A.}~\bibnamefont {Crook}}, \bibinfo {author} {\bibfnamefont
  {J.}~\bibnamefont {Bovaird}}, \bibinfo {author} {\bibfnamefont {J.~M.}\
  \bibnamefont {Kreikebaum}}, \bibinfo {author} {\bibfnamefont
  {N.}~\bibnamefont {Zobrist}}, \bibinfo {author} {\bibfnamefont
  {E.}~\bibnamefont {Jeffrey}}, \bibinfo {author} {\bibfnamefont
  {B.}~\bibnamefont {Ying}}, \bibinfo {author} {\bibfnamefont {A.}~\bibnamefont
  {Bengtsson}}, \bibinfo {author} {\bibfnamefont {H.-S.}\ \bibnamefont
  {Chang}}, \bibinfo {author} {\bibfnamefont {A.}~\bibnamefont {Dunsworth}},
  \bibinfo {author} {\bibfnamefont {J.}~\bibnamefont {Kelly}}, \bibinfo
  {author} {\bibfnamefont {Y.}~\bibnamefont {Zhang}}, \bibinfo {author}
  {\bibfnamefont {E.}~\bibnamefont {Forati}}, \bibinfo {author} {\bibfnamefont
  {R.}~\bibnamefont {Acharya}}, \bibinfo {author} {\bibfnamefont
  {J.}~\bibnamefont {Iveland}}, \bibinfo {author} {\bibfnamefont
  {W.}~\bibnamefont {Liu}}, \bibinfo {author} {\bibfnamefont {S.}~\bibnamefont
  {Kim}}, \bibinfo {author} {\bibfnamefont {B.}~\bibnamefont {Burkett}},
  \bibinfo {author} {\bibfnamefont {A.}~\bibnamefont {Megrant}}, \bibinfo
  {author} {\bibfnamefont {Y.}~\bibnamefont {Chen}}, \bibinfo {author}
  {\bibfnamefont {C.}~\bibnamefont {Neill}}, \bibinfo {author} {\bibfnamefont
  {D.}~\bibnamefont {Sank}}, \bibinfo {author} {\bibfnamefont {M.}~\bibnamefont
  {Devoret}},\ and\ \bibinfo {author} {\bibfnamefont {A.}~\bibnamefont
  {Opremcak}},\ }\bibfield  {title} {\bibinfo {title} {Resisting high-energy
  impact events through gap engineering in superconducting qubit arrays},\
  }\href {https://doi.org/10.1103/PhysRevLett.133.240601} {\bibfield  {journal}
  {\bibinfo  {journal} {Phys. Rev. Lett.}\ }\textbf {\bibinfo {volume} {133}},\
  \bibinfo {pages} {240601} (\bibinfo {year} {2024})}\BibitemShut {NoStop}%
\bibitem [{\citenamefont {Thorbeck}\ \emph {et~al.}(2023)\citenamefont
  {Thorbeck}, \citenamefont {Eddins}, \citenamefont {Lauer}, \citenamefont
  {McClure},\ and\ \citenamefont {Carroll}}]{Thorbeck_2023_TLS_impact_IBM}%
  \BibitemOpen
  \bibfield  {author} {\bibinfo {author} {\bibfnamefont {T.}~\bibnamefont
  {Thorbeck}}, \bibinfo {author} {\bibfnamefont {A.}~\bibnamefont {Eddins}},
  \bibinfo {author} {\bibfnamefont {I.}~\bibnamefont {Lauer}}, \bibinfo
  {author} {\bibfnamefont {D.~T.}\ \bibnamefont {McClure}},\ and\ \bibinfo
  {author} {\bibfnamefont {M.}~\bibnamefont {Carroll}},\ }\bibfield  {title}
  {\bibinfo {title} {Two-level-system dynamics in a superconducting qubit due
  to background ionizing radiation},\ }\bibfield  {journal} {\bibinfo
  {journal} {PRX Quantum}\ }\textbf {\bibinfo {volume} {4}},\ \href
  {https://doi.org/10.1103/prxquantum.4.020356} {10.1103/prxquantum.4.020356}
  (\bibinfo {year} {2023})\BibitemShut {NoStop}%
\bibitem [{\citenamefont {Barrett}\ \emph {et~al.}(2023)\citenamefont
  {Barrett}, \citenamefont {Karamlou}, \citenamefont {Muschinske},
  \citenamefont {Rosen}, \citenamefont {Braumüller}, \citenamefont {Das},
  \citenamefont {Kim}, \citenamefont {Niedzielski}, \citenamefont {Schuldt},
  \citenamefont {Serniak}, \citenamefont {Schwartz}, \citenamefont {Yoder},
  \citenamefont {Orlando}, \citenamefont {Gustavsson}, \citenamefont {Grover},\
  and\ \citenamefont {Oliver}}]{Barrett_2023_freq_calibration_error}%
  \BibitemOpen
  \bibfield  {author} {\bibinfo {author} {\bibfnamefont {C.~N.}\ \bibnamefont
  {Barrett}}, \bibinfo {author} {\bibfnamefont {A.~H.}\ \bibnamefont
  {Karamlou}}, \bibinfo {author} {\bibfnamefont {S.~E.}\ \bibnamefont
  {Muschinske}}, \bibinfo {author} {\bibfnamefont {I.~T.}\ \bibnamefont
  {Rosen}}, \bibinfo {author} {\bibfnamefont {J.}~\bibnamefont {Braumüller}},
  \bibinfo {author} {\bibfnamefont {R.}~\bibnamefont {Das}}, \bibinfo {author}
  {\bibfnamefont {D.~K.}\ \bibnamefont {Kim}}, \bibinfo {author} {\bibfnamefont
  {B.~M.}\ \bibnamefont {Niedzielski}}, \bibinfo {author} {\bibfnamefont
  {M.}~\bibnamefont {Schuldt}}, \bibinfo {author} {\bibfnamefont
  {K.}~\bibnamefont {Serniak}}, \bibinfo {author} {\bibfnamefont {M.~E.}\
  \bibnamefont {Schwartz}}, \bibinfo {author} {\bibfnamefont {J.~L.}\
  \bibnamefont {Yoder}}, \bibinfo {author} {\bibfnamefont {T.~P.}\ \bibnamefont
  {Orlando}}, \bibinfo {author} {\bibfnamefont {S.}~\bibnamefont {Gustavsson}},
  \bibinfo {author} {\bibfnamefont {J.~A.}\ \bibnamefont {Grover}},\ and\
  \bibinfo {author} {\bibfnamefont {W.~D.}\ \bibnamefont {Oliver}},\ }\bibfield
   {title} {\bibinfo {title} {Learning-based calibration of flux crosstalk in
  transmon qubit arrays},\ }\bibfield  {journal} {\bibinfo  {journal} {Physical
  Review Applied}\ }\textbf {\bibinfo {volume} {20}},\ \href
  {https://doi.org/10.1103/physrevapplied.20.024070}
  {10.1103/physrevapplied.20.024070} (\bibinfo {year} {2023})\BibitemShut
  {NoStop}%
\bibitem [{\citenamefont {Kanazawa}\ \emph {et~al.}(2023)\citenamefont
  {Kanazawa}, \citenamefont {Egger}, \citenamefont {Ben-Haim}, \citenamefont
  {Zhang}, \citenamefont {Shanks}, \citenamefont {Aleksandrowicz},\ and\
  \citenamefont {Wood}}]{Kanazawa2023_Qiskit_exp}%
  \BibitemOpen
  \bibfield  {author} {\bibinfo {author} {\bibfnamefont {N.}~\bibnamefont
  {Kanazawa}}, \bibinfo {author} {\bibfnamefont {D.~J.}\ \bibnamefont {Egger}},
  \bibinfo {author} {\bibfnamefont {Y.}~\bibnamefont {Ben-Haim}}, \bibinfo
  {author} {\bibfnamefont {H.}~\bibnamefont {Zhang}}, \bibinfo {author}
  {\bibfnamefont {W.~E.}\ \bibnamefont {Shanks}}, \bibinfo {author}
  {\bibfnamefont {G.}~\bibnamefont {Aleksandrowicz}},\ and\ \bibinfo {author}
  {\bibfnamefont {C.~J.}\ \bibnamefont {Wood}},\ }\bibfield  {title} {\bibinfo
  {title} {Qiskit experiments: A python package to characterize and calibrate
  quantum computers},\ }\href {https://doi.org/10.21105/joss.05329} {\bibfield
  {journal} {\bibinfo  {journal} {Journal of Open Source Software}\ }\textbf
  {\bibinfo {volume} {8}},\ \bibinfo {pages} {5329} (\bibinfo {year}
  {2023})}\BibitemShut {NoStop}%
\bibitem [{\citenamefont {de~Graaf}\ \emph {et~al.}(2018)\citenamefont
  {de~Graaf}, \citenamefont {Faoro}, \citenamefont {Burnett}, \citenamefont
  {Adamyan}, \citenamefont {Tzalenchuk}, \citenamefont {Kubatkin},
  \citenamefont {Lindström},\ and\ \citenamefont {Danilov}}]{de_Graaf_2018}%
  \BibitemOpen
  \bibfield  {author} {\bibinfo {author} {\bibfnamefont {S.~E.}\ \bibnamefont
  {de~Graaf}}, \bibinfo {author} {\bibfnamefont {L.}~\bibnamefont {Faoro}},
  \bibinfo {author} {\bibfnamefont {J.}~\bibnamefont {Burnett}}, \bibinfo
  {author} {\bibfnamefont {A.~A.}\ \bibnamefont {Adamyan}}, \bibinfo {author}
  {\bibfnamefont {A.~Y.}\ \bibnamefont {Tzalenchuk}}, \bibinfo {author}
  {\bibfnamefont {S.~E.}\ \bibnamefont {Kubatkin}}, \bibinfo {author}
  {\bibfnamefont {T.}~\bibnamefont {Lindström}},\ and\ \bibinfo {author}
  {\bibfnamefont {A.~V.}\ \bibnamefont {Danilov}},\ }\bibfield  {title}
  {\bibinfo {title} {Suppression of low-frequency charge noise in
  superconducting resonators by surface spin desorption},\ }\bibfield
  {journal} {\bibinfo  {journal} {Nature Communications}\ }\textbf {\bibinfo
  {volume} {9}},\ \href {https://doi.org/10.1038/s41467-018-03577-2}
  {10.1038/s41467-018-03577-2} (\bibinfo {year} {2018})\BibitemShut {NoStop}%
\bibitem [{\citenamefont {Koch}\ \emph {et~al.}(2007)\citenamefont {Koch},
  \citenamefont {Yu}, \citenamefont {Gambetta}, \citenamefont {Houck},
  \citenamefont {Schuster}, \citenamefont {Majer}, \citenamefont {Blais},
  \citenamefont {Devoret}, \citenamefont {Girvin},\ and\ \citenamefont
  {Schoelkopf}}]{Koch_2007}%
  \BibitemOpen
  \bibfield  {author} {\bibinfo {author} {\bibfnamefont {J.}~\bibnamefont
  {Koch}}, \bibinfo {author} {\bibfnamefont {T.~M.}\ \bibnamefont {Yu}},
  \bibinfo {author} {\bibfnamefont {J.}~\bibnamefont {Gambetta}}, \bibinfo
  {author} {\bibfnamefont {A.~A.}\ \bibnamefont {Houck}}, \bibinfo {author}
  {\bibfnamefont {D.~I.}\ \bibnamefont {Schuster}}, \bibinfo {author}
  {\bibfnamefont {J.}~\bibnamefont {Majer}}, \bibinfo {author} {\bibfnamefont
  {A.}~\bibnamefont {Blais}}, \bibinfo {author} {\bibfnamefont {M.~H.}\
  \bibnamefont {Devoret}}, \bibinfo {author} {\bibfnamefont {S.~M.}\
  \bibnamefont {Girvin}},\ and\ \bibinfo {author} {\bibfnamefont {R.~J.}\
  \bibnamefont {Schoelkopf}},\ }\bibfield  {title} {\bibinfo {title}
  {Charge-insensitive qubit design derived from the cooper pair box},\
  }\bibfield  {journal} {\bibinfo  {journal} {Physical Review A}\ }\textbf
  {\bibinfo {volume} {76}},\ \href {https://doi.org/10.1103/physreva.76.042319}
  {10.1103/physreva.76.042319} (\bibinfo {year} {2007})\BibitemShut {NoStop}%
\bibitem [{\citenamefont {Goelman}\ \emph {et~al.}(1989)\citenamefont
  {Goelman}, \citenamefont {Vega},\ and\ \citenamefont {Zax}}]{GOELMAN1989423}%
  \BibitemOpen
  \bibfield  {author} {\bibinfo {author} {\bibfnamefont {G.}~\bibnamefont
  {Goelman}}, \bibinfo {author} {\bibfnamefont {S.}~\bibnamefont {Vega}},\ and\
  \bibinfo {author} {\bibfnamefont {D.}~\bibnamefont {Zax}},\ }\bibfield
  {title} {\bibinfo {title} {Squared amplitude-modulated composite pulses},\
  }\href {https://doi.org/https://doi.org/10.1016/0022-2364(89)90077-2}
  {\bibfield  {journal} {\bibinfo  {journal} {Journal of Magnetic Resonance
  (1969)}\ }\textbf {\bibinfo {volume} {81}},\ \bibinfo {pages} {423} (\bibinfo
  {year} {1989})}\BibitemShut {NoStop}%
\bibitem [{\citenamefont {Viola}\ and\ \citenamefont
  {Lloyd}(1998)}]{Viola1998}%
  \BibitemOpen
  \bibfield  {author} {\bibinfo {author} {\bibfnamefont {L.}~\bibnamefont
  {Viola}}\ and\ \bibinfo {author} {\bibfnamefont {S.}~\bibnamefont {Lloyd}},\
  }\bibfield  {title} {\bibinfo {title} {Dynamical suppression of decoherence
  in two-state quantum systems},\ }\bibfield  {journal} {\bibinfo  {journal}
  {Physical Review A - Atomic, Molecular, and Optical Physics}\ }\textbf
  {\bibinfo {volume} {58}},\ \href {https://doi.org/10.1103/PhysRevA.58.2733}
  {10.1103/PhysRevA.58.2733} (\bibinfo {year} {1998})\BibitemShut {NoStop}%
\bibitem [{\citenamefont {Biercuk}\ \emph {et~al.}(2009)\citenamefont
  {Biercuk}, \citenamefont {Uys}, \citenamefont {VanDevender}, \citenamefont
  {Shiga}, \citenamefont {Itano},\ and\ \citenamefont
  {Bollinger}}]{Biercuk2009}%
  \BibitemOpen
  \bibfield  {author} {\bibinfo {author} {\bibfnamefont {M.~J.}\ \bibnamefont
  {Biercuk}}, \bibinfo {author} {\bibfnamefont {H.}~\bibnamefont {Uys}},
  \bibinfo {author} {\bibfnamefont {A.~P.}\ \bibnamefont {VanDevender}},
  \bibinfo {author} {\bibfnamefont {N.}~\bibnamefont {Shiga}}, \bibinfo
  {author} {\bibfnamefont {W.~M.}\ \bibnamefont {Itano}},\ and\ \bibinfo
  {author} {\bibfnamefont {J.~J.}\ \bibnamefont {Bollinger}},\ }\bibfield
  {title} {\bibinfo {title} {Optimized dynamical decoupling in a model quantum
  memory},\ }\bibfield  {journal} {\bibinfo  {journal} {Nature}\ }\textbf
  {\bibinfo {volume} {458}},\ \href {https://doi.org/10.1038/nature07951}
  {10.1038/nature07951} (\bibinfo {year} {2009})\BibitemShut {NoStop}%
\bibitem [{\citenamefont {Khodjasteh}\ and\ \citenamefont
  {Viola}(2009)}]{Khodjasteh2009}%
  \BibitemOpen
  \bibfield  {author} {\bibinfo {author} {\bibfnamefont {K.}~\bibnamefont
  {Khodjasteh}}\ and\ \bibinfo {author} {\bibfnamefont {L.}~\bibnamefont
  {Viola}},\ }\bibfield  {title} {\bibinfo {title} {Dynamically error-corrected
  gates for universal quantum computation},\ }\bibfield  {journal} {\bibinfo
  {journal} {Physical Review Letters}\ }\textbf {\bibinfo {volume} {102}},\
  \href {https://doi.org/10.1103/PhysRevLett.102.080501}
  {10.1103/PhysRevLett.102.080501} (\bibinfo {year} {2009})\BibitemShut
  {NoStop}%
\bibitem [{\citenamefont {Khodjasteh}\ \emph {et~al.}(2010)\citenamefont
  {Khodjasteh}, \citenamefont {Lidar},\ and\ \citenamefont
  {Viola}}]{Khodjasteh2010}%
  \BibitemOpen
  \bibfield  {author} {\bibinfo {author} {\bibfnamefont {K.}~\bibnamefont
  {Khodjasteh}}, \bibinfo {author} {\bibfnamefont {D.~A.}\ \bibnamefont
  {Lidar}},\ and\ \bibinfo {author} {\bibfnamefont {L.}~\bibnamefont {Viola}},\
  }\bibfield  {title} {\bibinfo {title} {Arbitrarily accurate dynamical control
  in open quantum systems},\ }\bibfield  {journal} {\bibinfo  {journal}
  {Physical Review Letters}\ }\textbf {\bibinfo {volume} {104}},\ \href
  {https://doi.org/10.1103/PhysRevLett.104.090501}
  {10.1103/PhysRevLett.104.090501} (\bibinfo {year} {2010})\BibitemShut
  {NoStop}%
\bibitem [{\citenamefont {Wang}\ \emph {et~al.}(2012)\citenamefont {Wang},
  \citenamefont {Bishop}, \citenamefont {Kestner}, \citenamefont {Barnes},
  \citenamefont {Sun},\ and\ \citenamefont {Sarma}}]{Wang2012}%
  \BibitemOpen
  \bibfield  {author} {\bibinfo {author} {\bibfnamefont {X.}~\bibnamefont
  {Wang}}, \bibinfo {author} {\bibfnamefont {L.~S.}\ \bibnamefont {Bishop}},
  \bibinfo {author} {\bibfnamefont {J.~P.}\ \bibnamefont {Kestner}}, \bibinfo
  {author} {\bibfnamefont {E.}~\bibnamefont {Barnes}}, \bibinfo {author}
  {\bibfnamefont {K.}~\bibnamefont {Sun}},\ and\ \bibinfo {author}
  {\bibfnamefont {D.}~\bibnamefont {Sarma}},\ }\bibfield  {title} {\bibinfo
  {title} {Composite pulses for robust universal control of singlet-triplet
  qubits},\ }\bibfield  {journal} {\bibinfo  {journal} {Nature Communications}\
  }\textbf {\bibinfo {volume} {3}},\ \href {https://doi.org/10.1038/ncomms2003}
  {10.1038/ncomms2003} (\bibinfo {year} {2012})\BibitemShut {NoStop}%
\bibitem [{\citenamefont {Sar}\ \emph {et~al.}(2012)\citenamefont {Sar},
  \citenamefont {Wang}, \citenamefont {Blok}, \citenamefont {Bernien},
  \citenamefont {Taminiau}, \citenamefont {Toyli}, \citenamefont {Lidar},
  \citenamefont {Awschalom}, \citenamefont {Hanson},\ and\ \citenamefont
  {Dobrovitski}}]{DCGGateDecoherence}%
  \BibitemOpen
  \bibfield  {author} {\bibinfo {author} {\bibfnamefont {T.~V.~D.}\
  \bibnamefont {Sar}}, \bibinfo {author} {\bibfnamefont {Z.~H.}\ \bibnamefont
  {Wang}}, \bibinfo {author} {\bibfnamefont {M.~S.}\ \bibnamefont {Blok}},
  \bibinfo {author} {\bibfnamefont {H.}~\bibnamefont {Bernien}}, \bibinfo
  {author} {\bibfnamefont {T.~H.}\ \bibnamefont {Taminiau}}, \bibinfo {author}
  {\bibfnamefont {D.~M.}\ \bibnamefont {Toyli}}, \bibinfo {author}
  {\bibfnamefont {D.~A.}\ \bibnamefont {Lidar}}, \bibinfo {author}
  {\bibfnamefont {D.~D.}\ \bibnamefont {Awschalom}}, \bibinfo {author}
  {\bibfnamefont {R.}~\bibnamefont {Hanson}},\ and\ \bibinfo {author}
  {\bibfnamefont {V.~V.}\ \bibnamefont {Dobrovitski}},\ }\bibfield  {title}
  {\bibinfo {title} {Decoherence-protected quantum gates for a hybrid
  solid-state spin register},\ }\bibfield  {journal} {\bibinfo  {journal}
  {Nature}\ }\textbf {\bibinfo {volume} {484}},\ \href
  {https://doi.org/10.1038/nature10900} {10.1038/nature10900} (\bibinfo {year}
  {2012})\BibitemShut {NoStop}%
\bibitem [{\citenamefont {Khodjasteh}\ \emph {et~al.}(2012)\citenamefont
  {Khodjasteh}, \citenamefont {Bluhm},\ and\ \citenamefont
  {Viola}}]{Khodjasteh2012}%
  \BibitemOpen
  \bibfield  {author} {\bibinfo {author} {\bibfnamefont {K.}~\bibnamefont
  {Khodjasteh}}, \bibinfo {author} {\bibfnamefont {H.}~\bibnamefont {Bluhm}},\
  and\ \bibinfo {author} {\bibfnamefont {L.}~\bibnamefont {Viola}},\ }\bibfield
   {title} {\bibinfo {title} {Automated synthesis of dynamically corrected
  quantum gates},\ }\bibfield  {journal} {\bibinfo  {journal} {Physical Review
  A - Atomic, Molecular, and Optical Physics}\ }\textbf {\bibinfo {volume}
  {86}},\ \href {https://doi.org/10.1103/PhysRevA.86.042329}
  {10.1103/PhysRevA.86.042329} (\bibinfo {year} {2012})\BibitemShut {NoStop}%
\bibitem [{\citenamefont {Kestner}\ \emph {et~al.}(2013)\citenamefont
  {Kestner}, \citenamefont {Wang}, \citenamefont {Bishop}, \citenamefont
  {Barnes},\ and\ \citenamefont {Sarma}}]{Kestner2013}%
  \BibitemOpen
  \bibfield  {author} {\bibinfo {author} {\bibfnamefont {J.~P.}\ \bibnamefont
  {Kestner}}, \bibinfo {author} {\bibfnamefont {X.}~\bibnamefont {Wang}},
  \bibinfo {author} {\bibfnamefont {L.~S.}\ \bibnamefont {Bishop}}, \bibinfo
  {author} {\bibfnamefont {E.}~\bibnamefont {Barnes}},\ and\ \bibinfo {author}
  {\bibfnamefont {S.~D.}\ \bibnamefont {Sarma}},\ }\bibfield  {title} {\bibinfo
  {title} {Noise-resistant control for a spin qubit array},\ }\bibfield
  {journal} {\bibinfo  {journal} {Physical Review Letters}\ }\textbf {\bibinfo
  {volume} {110}},\ \href {https://doi.org/10.1103/PhysRevLett.110.140502}
  {10.1103/PhysRevLett.110.140502} (\bibinfo {year} {2013})\BibitemShut
  {NoStop}%
\bibitem [{\citenamefont {Green}\ \emph {et~al.}(2013)\citenamefont {Green},
  \citenamefont {Sastrawan}, \citenamefont {Uys},\ and\ \citenamefont
  {Biercuk}}]{Green2013}%
  \BibitemOpen
  \bibfield  {author} {\bibinfo {author} {\bibfnamefont {T.~J.}\ \bibnamefont
  {Green}}, \bibinfo {author} {\bibfnamefont {J.}~\bibnamefont {Sastrawan}},
  \bibinfo {author} {\bibfnamefont {H.}~\bibnamefont {Uys}},\ and\ \bibinfo
  {author} {\bibfnamefont {M.~J.}\ \bibnamefont {Biercuk}},\ }\bibfield
  {title} {\bibinfo {title} {Arbitrary quantum control of qubits in the
  presence of universal noise},\ }\bibfield  {journal} {\bibinfo  {journal}
  {New Journal of Physics}\ }\textbf {\bibinfo {volume} {15}},\ \href
  {https://doi.org/10.1088/1367-2630/15/9/095004}
  {10.1088/1367-2630/15/9/095004} (\bibinfo {year} {2013})\BibitemShut
  {NoStop}%
\bibitem [{\citenamefont {Merrill}\ and\ \citenamefont
  {Brown}(2014)}]{DCG_Lie_KBrown}%
  \BibitemOpen
  \bibfield  {author} {\bibinfo {author} {\bibfnamefont {J.~T.}\ \bibnamefont
  {Merrill}}\ and\ \bibinfo {author} {\bibfnamefont {K.~R.}\ \bibnamefont
  {Brown}},\ }\bibinfo {title} {Progress in compensating pulse sequences for
  quantum computation},\ in\ \href
  {https://doi.org/https://doi.org/10.1002/9781118742631.ch10} {\emph {\bibinfo
  {booktitle} {Quantum Information and Computation for Chemistry}}}\ (\bibinfo
  {publisher} {John Wiley \& Sons, Ltd},\ \bibinfo {year} {2014})\ pp.\ \bibinfo
  {pages} {241--294}\BibitemShut {NoStop}%
\bibitem [{\citenamefont {Barnes}\ \emph {et~al.}(2015)\citenamefont {Barnes},
  \citenamefont {Wang},\ and\ \citenamefont {Sarma}}]{Barnes2015}%
  \BibitemOpen
  \bibfield  {author} {\bibinfo {author} {\bibfnamefont {E.}~\bibnamefont
  {Barnes}}, \bibinfo {author} {\bibfnamefont {X.}~\bibnamefont {Wang}},\ and\
  \bibinfo {author} {\bibfnamefont {S.~D.}\ \bibnamefont {Sarma}},\ }\bibfield
  {title} {\bibinfo {title} {Robust quantum control using smooth pulses and
  topological winding},\ }\bibfield  {journal} {\bibinfo  {journal} {Scientific
  Reports}\ }\textbf {\bibinfo {volume} {5}},\ \href
  {https://doi.org/10.1038/srep12685} {10.1038/srep12685} (\bibinfo {year}
  {2015})\BibitemShut {NoStop}%
\bibitem [{\citenamefont {Calderon-Vargas}\ and\ \citenamefont
  {Kestner}(2017)}]{DCGCNOT}%
  \BibitemOpen
  \bibfield  {author} {\bibinfo {author} {\bibfnamefont {F.~A.}\ \bibnamefont
  {Calderon-Vargas}}\ and\ \bibinfo {author} {\bibfnamefont {J.~P.}\
  \bibnamefont {Kestner}},\ }\bibfield  {title} {\bibinfo {title} {Dynamically
  correcting a cnot gate for any systematic logical error},\ }\bibfield
  {journal} {\bibinfo  {journal} {Physical Review Letters}\ }\textbf {\bibinfo
  {volume} {118}},\ \href {https://doi.org/10.1103/PhysRevLett.118.150502}
  {10.1103/PhysRevLett.118.150502} (\bibinfo {year} {2017})\BibitemShut
  {NoStop}%
\bibitem [{\citenamefont {Buterakos}\ \emph {et~al.}(2018)\citenamefont
  {Buterakos}, \citenamefont {Throckmorton},\ and\ \citenamefont
  {Sarma}}]{Buterakos2018}%
  \BibitemOpen
  \bibfield  {author} {\bibinfo {author} {\bibfnamefont {D.}~\bibnamefont
  {Buterakos}}, \bibinfo {author} {\bibfnamefont {R.~E.}\ \bibnamefont
  {Throckmorton}},\ and\ \bibinfo {author} {\bibfnamefont {S.~D.}\ \bibnamefont
  {Sarma}},\ }\bibfield  {title} {\bibinfo {title} {Crosstalk error correction
  through dynamical decoupling of single-qubit gates in capacitively coupled
  singlet-triplet semiconductor spin qubits},\ }\bibfield  {journal} {\bibinfo
  {journal} {Physical Review B}\ }\textbf {\bibinfo {volume} {97}},\ \href
  {https://doi.org/10.1103/PhysRevB.97.045431} {10.1103/PhysRevB.97.045431}
  (\bibinfo {year} {2018})\BibitemShut {NoStop}%
\bibitem [{\citenamefont {Güngördü}\ and\ \citenamefont
  {Kestner}(2022)}]{Kestner_2022_DCG_NN}%
  \BibitemOpen
  \bibfield  {author} {\bibinfo {author} {\bibfnamefont {U.}~\bibnamefont
  {Güngördü}}\ and\ \bibinfo {author} {\bibfnamefont {J.~P.}\ \bibnamefont
  {Kestner}},\ }\bibfield  {title} {\bibinfo {title} {Robust quantum gates
  using smooth pulses and physics-informed neural networks},\ }\bibfield
  {journal} {\bibinfo  {journal} {Physical Review Research}\ }\textbf {\bibinfo
  {volume} {4}},\ \href {https://doi.org/10.1103/physrevresearch.4.023155}
  {10.1103/physrevresearch.4.023155} (\bibinfo {year} {2022})\BibitemShut
  {NoStop}%
\bibitem [{\citenamefont {Güngördü}\ and\ \citenamefont
  {Kestner}(2018)}]{UtkanDCG2qub}%
  \BibitemOpen
  \bibfield  {author} {\bibinfo {author} {\bibfnamefont {U.}~\bibnamefont
  {Güngördü}}\ and\ \bibinfo {author} {\bibfnamefont {J.~P.}\ \bibnamefont
  {Kestner}},\ }\bibfield  {title} {\bibinfo {title} {Pulse sequence designed
  for robust c -phase gates in simos and si/sige double quantum dots},\
  }\bibfield  {journal} {\bibinfo  {journal} {Physical Review B}\ }\textbf
  {\bibinfo {volume} {98}},\ \href {https://doi.org/10.1103/PhysRevB.98.165301}
  {10.1103/PhysRevB.98.165301} (\bibinfo {year} {2018})\BibitemShut {NoStop}%
\bibitem [{\citenamefont {Kanaar}\ \emph {et~al.}(2021)\citenamefont {Kanaar},
  \citenamefont {Wolin}, \citenamefont {Güngördü},\ and\ \citenamefont
  {Kestner}}]{Kanaar2021}%
  \BibitemOpen
  \bibfield  {author} {\bibinfo {author} {\bibfnamefont {D.~W.}\ \bibnamefont
  {Kanaar}}, \bibinfo {author} {\bibfnamefont {S.}~\bibnamefont {Wolin}},
  \bibinfo {author} {\bibfnamefont {U.}~\bibnamefont {Güngördü}},\ and\
  \bibinfo {author} {\bibfnamefont {J.~P.}\ \bibnamefont {Kestner}},\
  }\bibfield  {title} {\bibinfo {title} {Single-tone pulse sequences and robust
  two-tone shaped pulses for three silicon spin qubits with always-on
  exchange},\ }\bibfield  {journal} {\bibinfo  {journal} {Physical Review B}\
  }\textbf {\bibinfo {volume} {103}},\ \href
  {https://doi.org/10.1103/PhysRevB.103.235314} {10.1103/PhysRevB.103.235314}
  (\bibinfo {year} {2021})\BibitemShut {NoStop}%
\bibitem [{\citenamefont {Carvalho}\ \emph {et~al.}(2021)\citenamefont
  {Carvalho}, \citenamefont {Ball}, \citenamefont {Biercuk}, \citenamefont
  {Hush},\ and\ \citenamefont {Thomsen}}]{Carvalho2021}%
  \BibitemOpen
  \bibfield  {author} {\bibinfo {author} {\bibfnamefont {A.~R.}\ \bibnamefont
  {Carvalho}}, \bibinfo {author} {\bibfnamefont {H.}~\bibnamefont {Ball}},
  \bibinfo {author} {\bibfnamefont {M.~J.}\ \bibnamefont {Biercuk}}, \bibinfo
  {author} {\bibfnamefont {M.~R.}\ \bibnamefont {Hush}},\ and\ \bibinfo
  {author} {\bibfnamefont {F.}~\bibnamefont {Thomsen}},\ }\bibfield  {title}
  {\bibinfo {title} {Error-robust quantum logic optimization using a cloud
  quantum computer interface},\ }\bibfield  {journal} {\bibinfo  {journal}
  {Physical Review Applied}\ }\textbf {\bibinfo {volume} {15}},\ \href
  {https://doi.org/10.1103/PhysRevApplied.15.064054}
  {10.1103/PhysRevApplied.15.064054} (\bibinfo {year} {2021})\BibitemShut
  {NoStop}%
\bibitem [{\citenamefont {Zeng}\ \emph {et~al.}(2019)\citenamefont {Zeng},
  \citenamefont {Yang}, \citenamefont {Dzurak},\ and\ \citenamefont
  {Barnes}}]{Zeng2019}%
  \BibitemOpen
  \bibfield  {author} {\bibinfo {author} {\bibfnamefont {J.}~\bibnamefont
  {Zeng}}, \bibinfo {author} {\bibfnamefont {C.~H.}\ \bibnamefont {Yang}},
  \bibinfo {author} {\bibfnamefont {A.~S.}\ \bibnamefont {Dzurak}},\ and\
  \bibinfo {author} {\bibfnamefont {E.}~\bibnamefont {Barnes}},\ }\bibfield
  {title} {\bibinfo {title} {Geometric formalism for constructing arbitrary
  single-qubit dynamically corrected gates},\ }\bibfield  {journal} {\bibinfo
  {journal} {Physical Review A}\ }\textbf {\bibinfo {volume} {99}},\ \href
  {https://doi.org/10.1103/PhysRevA.99.052321} {10.1103/PhysRevA.99.052321}
  (\bibinfo {year} {2019})\BibitemShut {NoStop}%
\bibitem [{\citenamefont {Dong}\ \emph {et~al.}(2021)\citenamefont {Dong},
  \citenamefont {Zhuang}, \citenamefont {Economou},\ and\ \citenamefont
  {Barnes}}]{Dong2021}%
  \BibitemOpen
  \bibfield  {author} {\bibinfo {author} {\bibfnamefont {W.}~\bibnamefont
  {Dong}}, \bibinfo {author} {\bibfnamefont {F.}~\bibnamefont {Zhuang}},
  \bibinfo {author} {\bibfnamefont {S.~E.}\ \bibnamefont {Economou}},\ and\
  \bibinfo {author} {\bibfnamefont {E.}~\bibnamefont {Barnes}},\ }\bibfield
  {title} {\bibinfo {title} {Doubly geometric quantum control},\ }\bibfield
  {journal} {\bibinfo  {journal} {PRX Quantum}\ }\textbf {\bibinfo {volume}
  {2}},\ \href {https://doi.org/10.1103/PRXQuantum.2.030333}
  {10.1103/PRXQuantum.2.030333} (\bibinfo {year} {2021})\BibitemShut {NoStop}%
\bibitem [{\citenamefont {Buterakos}\ \emph {et~al.}(2021)\citenamefont
  {Buterakos}, \citenamefont {Sarma},\ and\ \citenamefont
  {Barnes}}]{Buterakos2021}%
  \BibitemOpen
  \bibfield  {author} {\bibinfo {author} {\bibfnamefont {D.}~\bibnamefont
  {Buterakos}}, \bibinfo {author} {\bibfnamefont {S.~D.}\ \bibnamefont
  {Sarma}},\ and\ \bibinfo {author} {\bibfnamefont {E.}~\bibnamefont
  {Barnes}},\ }\bibfield  {title} {\bibinfo {title} {Geometrical formalism for
  dynamically corrected gates in multiqubit systems},\ }\bibfield  {journal}
  {\bibinfo  {journal} {PRX Quantum}\ }\textbf {\bibinfo {volume} {2}},\ \href
  {https://doi.org/10.1103/PRXQuantum.2.010341} {10.1103/PRXQuantum.2.010341}
  (\bibinfo {year} {2021})\BibitemShut {NoStop}%
\bibitem [{\citenamefont {Li}\ \emph {et~al.}(2021)\citenamefont {Li},
  \citenamefont {Calderon-Vargas}, \citenamefont {Zeng},\ and\ \citenamefont
  {Barnes}}]{Li2021}%
  \BibitemOpen
  \bibfield  {author} {\bibinfo {author} {\bibfnamefont {B.}~\bibnamefont
  {Li}}, \bibinfo {author} {\bibfnamefont {F.~A.}\ \bibnamefont
  {Calderon-Vargas}}, \bibinfo {author} {\bibfnamefont {J.}~\bibnamefont
  {Zeng}},\ and\ \bibinfo {author} {\bibfnamefont {E.}~\bibnamefont {Barnes}},\
  }\bibfield  {title} {\bibinfo {title} {Designing arbitrary single-axis
  rotations robust against perpendicular time-dependent noise},\ }\bibfield
  {journal} {\bibinfo  {journal} {New Journal of Physics}\ }\textbf {\bibinfo
  {volume} {23}},\ \href {https://doi.org/10.1088/1367-2630/ac22ea}
  {10.1088/1367-2630/ac22ea} (\bibinfo {year} {2021})\BibitemShut {NoStop}%
\bibitem [{\citenamefont {Barnes}\ \emph {et~al.}(2022)\citenamefont {Barnes},
  \citenamefont {Calderon-Vargas}, \citenamefont {Dong}, \citenamefont {Li},
  \citenamefont {Zeng},\ and\ \citenamefont {Zhuang}}]{Barnes2022}%
  \BibitemOpen
  \bibfield  {author} {\bibinfo {author} {\bibfnamefont {E.}~\bibnamefont
  {Barnes}}, \bibinfo {author} {\bibfnamefont {F.~A.}\ \bibnamefont
  {Calderon-Vargas}}, \bibinfo {author} {\bibfnamefont {W.}~\bibnamefont
  {Dong}}, \bibinfo {author} {\bibfnamefont {B.}~\bibnamefont {Li}}, \bibinfo
  {author} {\bibfnamefont {J.}~\bibnamefont {Zeng}},\ and\ \bibinfo {author}
  {\bibfnamefont {F.}~\bibnamefont {Zhuang}},\ }\href
  {https://doi.org/10.1088/2058-9565/ac4421} {\bibinfo {title} {Dynamically
  corrected gates from geometric space curves}} (\bibinfo {year}
  {2022})\BibitemShut {NoStop}%
\bibitem [{\citenamefont {Zhuang}\ \emph {et~al.}(2022)\citenamefont {Zhuang},
  \citenamefont {Zeng}, \citenamefont {Economou},\ and\ \citenamefont
  {Barnes}}]{Zhuang2022}%
  \BibitemOpen
  \bibfield  {author} {\bibinfo {author} {\bibfnamefont {F.}~\bibnamefont
  {Zhuang}}, \bibinfo {author} {\bibfnamefont {J.}~\bibnamefont {Zeng}},
  \bibinfo {author} {\bibfnamefont {S.~E.}\ \bibnamefont {Economou}},\ and\
  \bibinfo {author} {\bibfnamefont {E.}~\bibnamefont {Barnes}},\ }\bibfield
  {title} {\bibinfo {title} {Noise-resistant landau-zener sweeps from geometric
  curves},\ }\bibfield  {journal} {\bibinfo  {journal} {Quantum}\ }\textbf
  {\bibinfo {volume} {6}},\ \href {https://doi.org/10.22331/Q-2022-02-02-639}
  {10.22331/Q-2022-02-02-639} (\bibinfo {year} {2022})\BibitemShut {NoStop}%
\bibitem [{\citenamefont {Nelson}\ \emph {et~al.}(2023)\citenamefont {Nelson},
  \citenamefont {Piliouras}, \citenamefont {Connelly},\ and\ \citenamefont
  {Barnes}}]{Nelson2023}%
  \BibitemOpen
  \bibfield  {author} {\bibinfo {author} {\bibfnamefont {H.~T.}\ \bibnamefont
  {Nelson}}, \bibinfo {author} {\bibfnamefont {E.}~\bibnamefont {Piliouras}},
  \bibinfo {author} {\bibfnamefont {K.}~\bibnamefont {Connelly}},\ and\
  \bibinfo {author} {\bibfnamefont {E.}~\bibnamefont {Barnes}},\ }\bibfield
  {title} {\bibinfo {title} {Designing dynamically corrected gates robust to
  multiple noise sources using geometric space curves},\ }\bibfield  {journal}
  {\bibinfo  {journal} {Physical Review A}\ }\textbf {\bibinfo {volume}
  {108}},\ \href {https://doi.org/10.1103/PhysRevA.108.012407}
  {10.1103/PhysRevA.108.012407} (\bibinfo {year} {2023})\BibitemShut {NoStop}%
\bibitem [{\citenamefont {Piliouras}\ \emph {et~al.}(2025)\citenamefont
  {Piliouras}, \citenamefont {Lucarelli},\ and\ \citenamefont
  {Barnes}}]{piliouras2025}%
  \BibitemOpen
  \bibfield  {author} {\bibinfo {author} {\bibfnamefont {E.}~\bibnamefont
  {Piliouras}}, \bibinfo {author} {\bibfnamefont {D.}~\bibnamefont
  {Lucarelli}},\ and\ \bibinfo {author} {\bibfnamefont {E.}~\bibnamefont
  {Barnes}},\ }\href {https://arxiv.org/abs/2503.11492} {\bibinfo {title} {An
  automated geometric space curve approach for designing dynamically corrected
  gates}} (\bibinfo {year} {2025}),\ \Eprint {https://arxiv.org/abs/2503.11492}
  {arXiv:2503.11492 [quant-ph]} \BibitemShut {NoStop}%
\bibitem [{qur()}]{qurveros}%
  \BibitemOpen
  \href@noop {} {\bibinfo {title} {qurveros}},\ \bibinfo {howpublished}
  {\url{https://github.com/evpiliouras/qurveros}},\ \bibinfo {note} {accessed:
  2025-03-25}\BibitemShut {NoStop}%
\bibitem [{\citenamefont {Yi}\ \emph {et~al.}(2024)\citenamefont {Yi},
  \citenamefont {Hai}, \citenamefont {Luo}, \citenamefont {Chu}, \citenamefont
  {Zhang}, \citenamefont {Zhou}, \citenamefont {Song}, \citenamefont {Liu},
  \citenamefont {Yan}, \citenamefont {Deng}, \citenamefont {Chen},\ and\
  \citenamefont {Yu}}]{PhysRevLett.132.250604SCQC}%
  \BibitemOpen
  \bibfield  {author} {\bibinfo {author} {\bibfnamefont {K.}~\bibnamefont
  {Yi}}, \bibinfo {author} {\bibfnamefont {Y.-J.}\ \bibnamefont {Hai}},
  \bibinfo {author} {\bibfnamefont {K.}~\bibnamefont {Luo}}, \bibinfo {author}
  {\bibfnamefont {J.}~\bibnamefont {Chu}}, \bibinfo {author} {\bibfnamefont
  {L.}~\bibnamefont {Zhang}}, \bibinfo {author} {\bibfnamefont
  {Y.}~\bibnamefont {Zhou}}, \bibinfo {author} {\bibfnamefont {Y.}~\bibnamefont
  {Song}}, \bibinfo {author} {\bibfnamefont {S.}~\bibnamefont {Liu}}, \bibinfo
  {author} {\bibfnamefont {T.}~\bibnamefont {Yan}}, \bibinfo {author}
  {\bibfnamefont {X.-H.}\ \bibnamefont {Deng}}, \bibinfo {author}
  {\bibfnamefont {Y.}~\bibnamefont {Chen}},\ and\ \bibinfo {author}
  {\bibfnamefont {D.}~\bibnamefont {Yu}},\ }\bibfield  {title} {\bibinfo
  {title} {Robust quantum gates against correlated noise in integrated quantum
  chips},\ }\href {https://doi.org/10.1103/PhysRevLett.132.250604} {\bibfield
  {journal} {\bibinfo  {journal} {Phys. Rev. Lett.}\ }\textbf {\bibinfo
  {volume} {132}},\ \bibinfo {pages} {250604} (\bibinfo {year}
  {2024})}\BibitemShut {NoStop}%
\bibitem [{\citenamefont {Walelign}\ \emph {et~al.}(2024)\citenamefont
  {Walelign}, \citenamefont {Cai}, \citenamefont {Li}, \citenamefont {Barnes},\
  and\ \citenamefont {Nichol}}]{walelign2024dynamicallycorrectedgatessilicon}%
  \BibitemOpen
  \bibfield  {author} {\bibinfo {author} {\bibfnamefont {H.~Y.}\ \bibnamefont
  {Walelign}}, \bibinfo {author} {\bibfnamefont {X.}~\bibnamefont {Cai}},
  \bibinfo {author} {\bibfnamefont {B.}~\bibnamefont {Li}}, \bibinfo {author}
  {\bibfnamefont {E.}~\bibnamefont {Barnes}},\ and\ \bibinfo {author}
  {\bibfnamefont {J.~M.}\ \bibnamefont {Nichol}},\ }\href
  {https://arxiv.org/abs/2405.15148} {\bibinfo {title} {Dynamically corrected
  gates in silicon singlet-triplet spin qubits}} (\bibinfo {year} {2024}),\
  \Eprint {https://arxiv.org/abs/2405.15148} {arXiv:2405.15148 [quant-ph]}
  \BibitemShut {NoStop}%
\bibitem [{\citenamefont {Poggi}\ \emph {et~al.}(2024)\citenamefont {Poggi},
  \citenamefont {De~Chiara}, \citenamefont {Campbell},\ and\ \citenamefont
  {Kiely}}]{Poggi_2024}%
  \BibitemOpen
  \bibfield  {author} {\bibinfo {author} {\bibfnamefont {P.~M.}\ \bibnamefont
  {Poggi}}, \bibinfo {author} {\bibfnamefont {G.}~\bibnamefont {De~Chiara}},
  \bibinfo {author} {\bibfnamefont {S.}~\bibnamefont {Campbell}},\ and\
  \bibinfo {author} {\bibfnamefont {A.}~\bibnamefont {Kiely}},\ }\bibfield
  {title} {\bibinfo {title} {Universally robust quantum control},\ }\bibfield
  {journal} {\bibinfo  {journal} {Physical Review Letters}\ }\textbf {\bibinfo
  {volume} {132}},\ \href {https://doi.org/10.1103/physrevlett.132.193801}
  {10.1103/physrevlett.132.193801} (\bibinfo {year} {2024})\BibitemShut
  {NoStop}%
\bibitem [{\citenamefont {Khaneja}\ \emph {et~al.}(2005)\citenamefont
  {Khaneja}, \citenamefont {Reiss}, \citenamefont {Kehlet}, \citenamefont
  {Schulte-Herbrüggen},\ and\ \citenamefont {Glaser}}]{Khaneja2005GRAPE}%
  \BibitemOpen
  \bibfield  {author} {\bibinfo {author} {\bibfnamefont {N.}~\bibnamefont
  {Khaneja}}, \bibinfo {author} {\bibfnamefont {T.}~\bibnamefont {Reiss}},
  \bibinfo {author} {\bibfnamefont {C.}~\bibnamefont {Kehlet}}, \bibinfo
  {author} {\bibfnamefont {T.}~\bibnamefont {Schulte-Herbrüggen}},\ and\
  \bibinfo {author} {\bibfnamefont {S.~J.}\ \bibnamefont {Glaser}},\ }\bibfield
   {title} {\bibinfo {title} {Optimal control of coupled spin dynamics: Design
  of nmr pulse sequences by gradient ascent algorithms},\ }\bibfield  {journal}
  {\bibinfo  {journal} {Journal of Magnetic Resonance}\ }\textbf {\bibinfo
  {volume} {172}},\ \href {https://doi.org/10.1016/j.jmr.2004.11.004}
  {10.1016/j.jmr.2004.11.004} (\bibinfo {year} {2005})\BibitemShut {NoStop}%
\bibitem [{\citenamefont {Doria}\ \emph {et~al.}(2011)\citenamefont {Doria},
  \citenamefont {Calarco},\ and\ \citenamefont {Montangero}}]{Doria_2011_CRAB}%
  \BibitemOpen
  \bibfield  {author} {\bibinfo {author} {\bibfnamefont {P.}~\bibnamefont
  {Doria}}, \bibinfo {author} {\bibfnamefont {T.}~\bibnamefont {Calarco}},\
  and\ \bibinfo {author} {\bibfnamefont {S.}~\bibnamefont {Montangero}},\
  }\bibfield  {title} {\bibinfo {title} {Optimal control technique for
  many-body quantum dynamics},\ }\href
  {https://doi.org/10.1103/PhysRevLett.106.190501} {\bibfield  {journal}
  {\bibinfo  {journal} {Phys. Rev. Lett.}\ }\textbf {\bibinfo {volume} {106}},\
  \bibinfo {pages} {190501} (\bibinfo {year} {2011})}\BibitemShut {NoStop}%
\bibitem [{\citenamefont {Jones}\ \emph {et~al.}(2012)\citenamefont {Jones},
  \citenamefont {Ladd},\ and\ \citenamefont {Fong}}]{Jones2012CORPSE}%
  \BibitemOpen
  \bibfield  {author} {\bibinfo {author} {\bibfnamefont {N.~C.}\ \bibnamefont
  {Jones}}, \bibinfo {author} {\bibfnamefont {T.~D.}\ \bibnamefont {Ladd}},\
  and\ \bibinfo {author} {\bibfnamefont {B.~H.}\ \bibnamefont {Fong}},\
  }\bibfield  {title} {\bibinfo {title} {Dynamical decoupling of a qubit with
  always-on control fields},\ }\bibfield  {journal} {\bibinfo  {journal} {New
  Journal of Physics}\ }\textbf {\bibinfo {volume} {14}},\ \href
  {https://doi.org/10.1088/1367-2630/14/9/093045}
  {10.1088/1367-2630/14/9/093045} (\bibinfo {year} {2012})\BibitemShut
  {NoStop}%
\bibitem [{\citenamefont {Machnes}\ \emph {et~al.}(2018)\citenamefont
  {Machnes}, \citenamefont {Ass\'emat}, \citenamefont {Tannor},\ and\
  \citenamefont {Wilhelm}}]{GOAT_PhysRevLett.120.150401}%
  \BibitemOpen
  \bibfield  {author} {\bibinfo {author} {\bibfnamefont {S.}~\bibnamefont
  {Machnes}}, \bibinfo {author} {\bibfnamefont {E.}~\bibnamefont {Ass\'emat}},
  \bibinfo {author} {\bibfnamefont {D.}~\bibnamefont {Tannor}},\ and\ \bibinfo
  {author} {\bibfnamefont {F.~K.}\ \bibnamefont {Wilhelm}},\ }\bibfield
  {title} {\bibinfo {title} {Tunable, flexible, and efficient optimization of
  control pulses for practical qubits},\ }\href
  {https://doi.org/10.1103/PhysRevLett.120.150401} {\bibfield  {journal}
  {\bibinfo  {journal} {Phys. Rev. Lett.}\ }\textbf {\bibinfo {volume} {120}},\
  \bibinfo {pages} {150401} (\bibinfo {year} {2018})}\BibitemShut {NoStop}%
\bibitem [{\citenamefont {Hansen}\ \emph {et~al.}(2022)\citenamefont {Hansen},
  \citenamefont {Seedhouse}, \citenamefont {Chan}, \citenamefont {Hudson},
  \citenamefont {Itoh}, \citenamefont {Laucht}, \citenamefont {Saraiva},
  \citenamefont {Yang},\ and\ \citenamefont {Dzurak}}]{Hansen_APR2022}%
  \BibitemOpen
  \bibfield  {author} {\bibinfo {author} {\bibfnamefont {I.}~\bibnamefont
  {Hansen}}, \bibinfo {author} {\bibfnamefont {A.~E.}\ \bibnamefont
  {Seedhouse}}, \bibinfo {author} {\bibfnamefont {K.~W.}\ \bibnamefont {Chan}},
  \bibinfo {author} {\bibfnamefont {F.~E.}\ \bibnamefont {Hudson}}, \bibinfo
  {author} {\bibfnamefont {K.~M.}\ \bibnamefont {Itoh}}, \bibinfo {author}
  {\bibfnamefont {A.}~\bibnamefont {Laucht}}, \bibinfo {author} {\bibfnamefont
  {A.}~\bibnamefont {Saraiva}}, \bibinfo {author} {\bibfnamefont {C.~H.}\
  \bibnamefont {Yang}},\ and\ \bibinfo {author} {\bibfnamefont {A.~S.}\
  \bibnamefont {Dzurak}},\ }\bibfield  {title} {\bibinfo {title}
  {Implementation of an advanced dressing protocol for global qubit control in
  silicon},\ }\href {https://doi.org/10.1063/5.0096467} {\bibfield  {journal}
  {\bibinfo  {journal} {Applied Physics Reviews}\ }\textbf {\bibinfo {volume}
  {9}},\ \bibinfo {pages} {031409} (\bibinfo {year} {2022})},\ \Eprint
  {https://arxiv.org/abs/https://pubs.aip.org/aip/apr/article-pdf/doi/10.1063/5.0096467/19807645/031409\_1\_online.pdf}
  {https://pubs.aip.org/aip/apr/article-pdf/doi/10.1063/5.0096467/19807645/031409\_1\_online.pdf}
  \BibitemShut {NoStop}%
\bibitem [{\citenamefont {McKay}\ \emph {et~al.}(2017)\citenamefont {McKay},
  \citenamefont {Wood}, \citenamefont {Sheldon}, \citenamefont {Chow},\ and\
  \citenamefont {Gambetta}}]{McKay_2017}%
  \BibitemOpen
  \bibfield  {author} {\bibinfo {author} {\bibfnamefont {D.~C.}\ \bibnamefont
  {McKay}}, \bibinfo {author} {\bibfnamefont {C.~J.}\ \bibnamefont {Wood}},
  \bibinfo {author} {\bibfnamefont {S.}~\bibnamefont {Sheldon}}, \bibinfo
  {author} {\bibfnamefont {J.~M.}\ \bibnamefont {Chow}},\ and\ \bibinfo
  {author} {\bibfnamefont {J.~M.}\ \bibnamefont {Gambetta}},\ }\bibfield
  {title} {\bibinfo {title} {Efficient z-gates for quantum computing},\
  }\bibfield  {journal} {\bibinfo  {journal} {Physical Review A}\ }\textbf
  {\bibinfo {volume} {96}},\ \href {https://doi.org/10.1103/physreva.96.022330}
  {10.1103/physreva.96.022330} (\bibinfo {year} {2017})\BibitemShut {NoStop}%
\bibitem [{\citenamefont {Sundaresan}\ \emph {et~al.}(2020)\citenamefont
  {Sundaresan}, \citenamefont {Lauer}, \citenamefont {Pritchett}, \citenamefont
  {Magesan}, \citenamefont {Jurcevic},\ and\ \citenamefont
  {Gambetta}}]{coherence_1_Sundaresan_2020}%
  \BibitemOpen
  \bibfield  {author} {\bibinfo {author} {\bibfnamefont {N.}~\bibnamefont
  {Sundaresan}}, \bibinfo {author} {\bibfnamefont {I.}~\bibnamefont {Lauer}},
  \bibinfo {author} {\bibfnamefont {E.}~\bibnamefont {Pritchett}}, \bibinfo
  {author} {\bibfnamefont {E.}~\bibnamefont {Magesan}}, \bibinfo {author}
  {\bibfnamefont {P.}~\bibnamefont {Jurcevic}},\ and\ \bibinfo {author}
  {\bibfnamefont {J.~M.}\ \bibnamefont {Gambetta}},\ }\bibfield  {title}
  {\bibinfo {title} {Reducing unitary and spectator errors in cross resonance
  with optimized rotary echoes},\ }\bibfield  {journal} {\bibinfo  {journal}
  {PRX Quantum}\ }\textbf {\bibinfo {volume} {1}},\ \href
  {https://doi.org/10.1103/prxquantum.1.020318} {10.1103/prxquantum.1.020318}
  (\bibinfo {year} {2020})\BibitemShut {NoStop}%
\bibitem [{\citenamefont {Garion}\ \emph {et~al.}(2021)\citenamefont {Garion},
  \citenamefont {Kanazawa}, \citenamefont {Landa}, \citenamefont {McKay},
  \citenamefont {Sheldon}, \citenamefont {Cross},\ and\ \citenamefont
  {Wood}}]{coherence_2_Garion_2021}%
  \BibitemOpen
  \bibfield  {author} {\bibinfo {author} {\bibfnamefont {S.}~\bibnamefont
  {Garion}}, \bibinfo {author} {\bibfnamefont {N.}~\bibnamefont {Kanazawa}},
  \bibinfo {author} {\bibfnamefont {H.}~\bibnamefont {Landa}}, \bibinfo
  {author} {\bibfnamefont {D.~C.}\ \bibnamefont {McKay}}, \bibinfo {author}
  {\bibfnamefont {S.}~\bibnamefont {Sheldon}}, \bibinfo {author} {\bibfnamefont
  {A.~W.}\ \bibnamefont {Cross}},\ and\ \bibinfo {author} {\bibfnamefont
  {C.~J.}\ \bibnamefont {Wood}},\ }\bibfield  {title} {\bibinfo {title}
  {Experimental implementation of non-clifford interleaved randomized
  benchmarking with a controlled-s gate},\ }\bibfield  {journal} {\bibinfo
  {journal} {Physical Review Research}\ }\textbf {\bibinfo {volume} {3}},\
  \href {https://doi.org/10.1103/physrevresearch.3.013204}
  {10.1103/physrevresearch.3.013204} (\bibinfo {year} {2021})\BibitemShut
  {NoStop}%
\bibitem [{\citenamefont {Wei}\ \emph {et~al.}(2024)\citenamefont {Wei},
  \citenamefont {Pritchett}, \citenamefont {Zajac}, \citenamefont {McKay},\
  and\ \citenamefont {Merkel}}]{coherence_3_Wei_2024}%
  \BibitemOpen
  \bibfield  {author} {\bibinfo {author} {\bibfnamefont {K.~X.}\ \bibnamefont
  {Wei}}, \bibinfo {author} {\bibfnamefont {E.}~\bibnamefont {Pritchett}},
  \bibinfo {author} {\bibfnamefont {D.~M.}\ \bibnamefont {Zajac}}, \bibinfo
  {author} {\bibfnamefont {D.~C.}\ \bibnamefont {McKay}},\ and\ \bibinfo
  {author} {\bibfnamefont {S.}~\bibnamefont {Merkel}},\ }\bibfield  {title}
  {\bibinfo {title} {Characterizing non-markovian off-resonant errors in
  quantum gates},\ }\bibfield  {journal} {\bibinfo  {journal} {Physical Review
  Applied}\ }\textbf {\bibinfo {volume} {21}},\ \href
  {https://doi.org/10.1103/physrevapplied.21.024018}
  {10.1103/physrevapplied.21.024018} (\bibinfo {year} {2024})\BibitemShut
  {NoStop}%
\bibitem [{\citenamefont {Alexander}\ \emph {et~al.}(2020)\citenamefont
  {Alexander}, \citenamefont {Kanazawa}, \citenamefont {Egger}, \citenamefont
  {Capelluto}, \citenamefont {Wood}, \citenamefont {Javadi-Abhari},\ and\
  \citenamefont {C~McKay}}]{Alexander_2020_Qsikit_pulse}%
  \BibitemOpen
  \bibfield  {author} {\bibinfo {author} {\bibfnamefont {T.}~\bibnamefont
  {Alexander}}, \bibinfo {author} {\bibfnamefont {N.}~\bibnamefont {Kanazawa}},
  \bibinfo {author} {\bibfnamefont {D.~J.}\ \bibnamefont {Egger}}, \bibinfo
  {author} {\bibfnamefont {L.}~\bibnamefont {Capelluto}}, \bibinfo {author}
  {\bibfnamefont {C.~J.}\ \bibnamefont {Wood}}, \bibinfo {author}
  {\bibfnamefont {A.}~\bibnamefont {Javadi-Abhari}},\ and\ \bibinfo {author}
  {\bibfnamefont {D.}~\bibnamefont {C~McKay}},\ }\bibfield  {title} {\bibinfo
  {title} {Qiskit pulse: programming quantum computers through the cloud with
  pulses},\ }\href {https://doi.org/10.1088/2058-9565/aba404} {\bibfield
  {journal} {\bibinfo  {journal} {Quantum Science and Technology}\ }\textbf
  {\bibinfo {volume} {5}},\ \bibinfo {pages} {044006} (\bibinfo {year}
  {2020})}\BibitemShut {NoStop}%
\bibitem [{\citenamefont {Motzoi}\ \emph {et~al.}(2009)\citenamefont {Motzoi},
  \citenamefont {Gambetta}, \citenamefont {Rebentrost},\ and\ \citenamefont
  {Wilhelm}}]{IBM_drag_2009}%
  \BibitemOpen
  \bibfield  {author} {\bibinfo {author} {\bibfnamefont {F.}~\bibnamefont
  {Motzoi}}, \bibinfo {author} {\bibfnamefont {J.~M.}\ \bibnamefont
  {Gambetta}}, \bibinfo {author} {\bibfnamefont {P.}~\bibnamefont
  {Rebentrost}},\ and\ \bibinfo {author} {\bibfnamefont {F.~K.}\ \bibnamefont
  {Wilhelm}},\ }\bibfield  {title} {\bibinfo {title} {Simple pulses for
  elimination of leakage in weakly nonlinear qubits},\ }\href
  {https://doi.org/10.1103/PhysRevLett.103.110501} {\bibfield  {journal}
  {\bibinfo  {journal} {Phys. Rev. Lett.}\ }\textbf {\bibinfo {volume} {103}},\
  \bibinfo {pages} {110501} (\bibinfo {year} {2009})}\BibitemShut {NoStop}%
\bibitem [{\citenamefont {Gambetta}\ \emph {et~al.}(2011)\citenamefont
  {Gambetta}, \citenamefont {Motzoi}, \citenamefont {Merkel},\ and\
  \citenamefont {Wilhelm}}]{IBM_drag_2011}%
  \BibitemOpen
  \bibfield  {author} {\bibinfo {author} {\bibfnamefont {J.~M.}\ \bibnamefont
  {Gambetta}}, \bibinfo {author} {\bibfnamefont {F.}~\bibnamefont {Motzoi}},
  \bibinfo {author} {\bibfnamefont {S.~T.}\ \bibnamefont {Merkel}},\ and\
  \bibinfo {author} {\bibfnamefont {F.~K.}\ \bibnamefont {Wilhelm}},\
  }\bibfield  {title} {\bibinfo {title} {Analytic control methods for
  high-fidelity unitary operations in a weakly nonlinear oscillator},\ }\href
  {https://doi.org/10.1103/PhysRevA.83.012308} {\bibfield  {journal} {\bibinfo
  {journal} {Phys. Rev. A}\ }\textbf {\bibinfo {volume} {83}},\ \bibinfo
  {pages} {012308} (\bibinfo {year} {2011})}\BibitemShut {NoStop}%
\bibitem [{\citenamefont {Zeng}\ and\ \citenamefont
  {Barnes}(2018)}]{Zeng_PRA2018}%
  \BibitemOpen
  \bibfield  {author} {\bibinfo {author} {\bibfnamefont {J.}~\bibnamefont
  {Zeng}}\ and\ \bibinfo {author} {\bibfnamefont {E.}~\bibnamefont {Barnes}},\
  }\bibfield  {title} {\bibinfo {title} {Fastest pulses that implement
  dynamically corrected single-qubit phase gates},\ }\href
  {https://doi.org/10.1103/PhysRevA.98.012301} {\bibfield  {journal} {\bibinfo
  {journal} {Phys. Rev. A}\ }\textbf {\bibinfo {volume} {98}},\ \bibinfo
  {pages} {012301} (\bibinfo {year} {2018})}\BibitemShut {NoStop}%
\end{thebibliography}%

\appendix

\section{A brief theoretical overview of the Space Curve Quantum Control (SCQC) formalism and the B\'ezier Ansatz for Robust Quantum (BARQ) control}
\label{app::theory_BARQ}

First we need to establish a Hamiltonian to take as our starting point. We start by considering the single qubit Hamiltonian that approximates an IBM device \cite{Alexander_2020_Qsikit_pulse} in the lab frame within a noiseless environment.
\begin{equation}
    \label{eqn:H_lab}
    H_{lab}(t)=-\frac{\omega_q}{2} \sigma_z
    +{\Omega(t)}\cos \left[\omega_d t-\Phi(t)\right] \sigma_x,
\end{equation}
where {$\sigma_x,\sigma_y,\sigma_z$} are the Pauli matrices, $\Omega(t)$ is the Rabi rate that follows from the magnitude of the modulated driving field, $\Phi(t)$ is its phase, $\omega_q$ is the qubit frequency for the $0 \rightarrow 1$ transition and $\omega_d$ is the carrier frequency with which the driving field oscillates. By taking this noiseless lab Hamiltonian to the drive frame defined by $U_d = e^{i \frac{\omega_dt}{2}\sigma_z}$ and applying the rotating wave approximation (RWA) to polish out the highly oscillatory terms, we arrive at the generic, noiseless, effective Hamiltonian
\begin{equation}
    \label{eqn:H_0_Appendix}
    H_{0}(t)=\frac{\Delta}{2} \sigma_z
    +\frac{\Omega(t)}{2}\left[\cos \Phi(t) \sigma_x+\sin \Phi(t) \sigma_y\right],
\end{equation}
where $\Delta$ is the detuning defined as $\Delta = \omega_d -\omega_q$. 
This $H_{0}$ Hamiltonian is the starting point for the SCQC analysis to be used in this work, it is also a good approximation for effective Hamiltonians on many platforms and is accordingly suitable for platform agnostic schemes.

Now that the noiseless Hamiltonian has been established we can introduce two environmental noise terms into the Hamiltonian. Errors can conveniently be modeled as additive longitudinal errors $\delta_z$ in $\Delta$ and multiplicative transverse errors $\epsilon$ in $\Omega(t)$, this gives the noisy Hamiltonian 
\begin{equation}
\label{eqn:H_noise}
\begin{aligned}
\begin{gathered}
\Tilde{H}_0(t)= \frac{\Omega(t)(1+\epsilon)}{2}(\cos \Phi(t) \sigma_x + \sin \Phi(t) \sigma_y)+ \\
\frac{\Delta(t)+\delta_z}{2} \sigma_z,
\end{gathered}
\end{aligned}
\end{equation}

We will also consider that the noise parameters are quasi-static noise terms, which is a viable assumption on many realistic systems. The total evolution is given by $U = U_0U_{\text{noise}}$. All the theoretical derivations discussed in this Appendix are covered extensively in Refs.\cite{Zeng2019,Nelson2023}. Following these references we can move into the noiseless dynamics frame defined by $U_0$, i.e., the unitary evolution generated by $H_0$ in Eq.~\eqref{eqn:H_0_Appendix}. The Magnus expansion is then used to expand $U_{\text{noise}}$ to first-order in the noisy parameters $(\epsilon,\delta_z)$ yielding 
\begin{equation}
\label{eqn:magnus_firstOrder_1}
U_{\text{noise}}(t) \approx e^{-i \Pi_1(t)}
\end{equation}
where 
\begin{equation}
\label{eqn:magnus_firstOrder_2}
\begin{aligned}
\Pi_1(t)=\int_0^t d t^{\prime} H_I\left(t^{\prime}\right)=\frac{\delta_z}{2} \int_0^t d t^{\prime} U_0^{\dagger}\left(t^{\prime}\right) \sigma_z U_0\left(t^{\prime}\right) + \\
\frac{\epsilon}{2} \int_0^t d t^{\prime} U_0^{\dagger}\left(t^{\prime}\right) \Omega\left(t^{\prime}\right)\left[\cos \Phi\left(t^{\prime}\right) \sigma_x+\sin \Phi\left(t^{\prime}\right) \sigma_y\right]U_0\left(t^{\prime}\right)
\end{aligned}
\end{equation}

Accordingly, we can suppress the errors induced by noise $(\epsilon,\delta_z)$ in $U_{\text{noise}}$ at the gate time $T_g$, which marks the end of the evolution, through the following conditions:
\begin{align}
    \delta_z:\int_0^{T_g} dt& \, U_0^\dagger \sigma_z U_0 = 0 ,\label{eqn:dephasing_robust_cond_appx}\\
     \epsilon:\int_0^{T_g} dt&\, U_0^\dagger\, \Omega(t)\left[\cos \Phi(t) \sigma_x+\sin \Phi(t) \sigma_y\right]U_0 =0 \label{eqn:amp_roubst_cond_appx}
\end{align}
which ensure $\Pi_1$ vanishes, and therefore $U_{\text{noise}}(T_g)\approx I$.

The relation to differential geometry can be established by using  Eq.~\eqref{eqn:dephasing_robust_cond_appx} to inspire the definition of an \textit{error curve} as per Ref. \cite{Zeng2019} giving rise to the relation
\begin{align}
    \int_0^{t} dt' \, U_0^\dagger \sigma_z U_0 = \vec r (t) \cdot \sigvec, \label{scqc_main_eq}
\end{align}
where $\vec r(t)$ is the position vector of a curve and $\sigvec = \begin{bmatrix}
    \sigma_x & \sigma_y & \sigma_z
\end{bmatrix}^T$.
This essentially maps the dephasing robustness condition in Eq.~\eqref{eqn:dephasing_robust_cond_appx} to the analytic geometric condition 
\begin{equation}
\label{eqn:closed_curve_geometry_app}
    \vec r(T_g)=\vec r(0),
\end{equation}
which is the closed-curve condition. Through this definition, SCQC basically associates dephasing robust controls with closed loops in 3D where time is equivalent to the length along the curve.

In order to fully characterize the quantum evolution in terms of the space curve, we can differentiate Eq.~\eqref{scqc_main_eq} and attach at each point along the curve, the Frenet-Serret (FS) frame, which is constructed from the vectors $\vec T,\vec N, \vec B$ called tangent, normal and binormal respectively. The tangent vector is defined as $\dot{ \vec r}= \vec T$, and the two orthogonal vectors as $\vec N = \dot{\vec T}/||\dot{\vec T}||_2$ with $\vec B = \vec T \times \vec N$. When the action of $U_0$ is understood as a quantum channel, the elements of the Pauli Transfer Matrix (PTM) are expressed in terms of the FS vectors \cite{piliouras2025}. The SCQC control waveforms that implement the Hamiltonian in Eq.~\eqref{eqn:H_0_Appendix}, in a noise robust fashion, can now be found through the relations
\begin{align}
\label{eqn:curvature_to_Omega}
    \Omega(t) &= \dot {\vec T} \cdot \vec N = \kappa(t),\\
    \dot \Phi(t) - \Delta &= \dot {\vec N} \cdot \vec B = \tau(t), 
\end{align}
where $\kappa(t), \tau(t)$ are the curvature and torsion functions respectively. The multiplicative error in the envelope is suppressed by requiring
\begin{align}
\label{eqn:zero_area_geometry_app}
    \int_0^{T_g} dt \, \vec{T} \times \dot{\vec T} = \vec 0,
\end{align}
which geometrically corresponds to requiring the tangent vector to trace zero (oriented) area \cite{Nelson2023}. 

As discussed in the main text, it is a geometrical fact that there is immense freedom in finding admissible control waveforms that satisfy both Eq.~\eqref{eqn:dephasing_robust_cond_appx} and Eq.~\eqref{eqn:amp_roubst_cond_appx}. This motivated the creation of BARQ~\cite{piliouras2025}, a control method that provides the toolkit to automate the creation of experimentally friendly and robust gates guided by the SCQC formalism.

In BARQ, the position vector $\vec r(x(t))$ is parameterized using a set of control points $\vec w_i$ and a set of basis functions $g_{i,n}(x(t))$ resulting in 
\begin{align}
    \vec r(x(t)) = \sum_{j=0}^{n} \vec w_j g_{j,n}(x(t)),\qquad x(t) \in [0,1], \label{bezier_curve_eq}
\end{align}
where
\begin{align}
    g_{j,n}(x) = \begin{pmatrix}
        n \\ j
    \end{pmatrix} x^j(1- x)^{n-j}
\end{align}
is the Bernstein basis. The function $x(t)$ is a monotonic function that ensures $||\dot {\vec r}(x(t))||=1, \, \forall t$ so that time is measured as the length along the curve. The control points are responsible for both fixing the target operation $U_0(T_g)$ and enforcing the robustness conditions. An important feature of BARQ is that the gate is always fixed for unit fidelity and any source of infidelity is purely attributed to the experimental implementation of the pulse. In that sense, the curve optimization (which corresponds to translating the control points based on a cost function $J$) takes place in a trade-off free subspace of parameters. The optimization is performed using the Python package \texttt{qurveros}~\cite{qurveros}.

\end{document}